\documentclass{aa}  

\usepackage{soul}
\usepackage{cancel}
\usepackage{graphicx}	
\usepackage{amsmath}	
\usepackage{mathtools,amssymb}
\usepackage{soul}
\usepackage{color}
\usepackage{natbib}
\usepackage{url}
\usepackage{placeins}
\usepackage{xfrac}
\usepackage{comment}
\usepackage[breaklinks=true]{hyperref} 
\usepackage{amssymb}
\usepackage{booktabs}
\usepackage{soul}
\usepackage{tabularx}
\usepackage{verbatim}
\usepackage{orcidlink}
\usepackage{multirow}
\usepackage{subcaption}
\usepackage{graphicx}
\usepackage{txfonts}

\usepackage{soul,xcolor}
\usepackage[normalem]{ulem}  

\definecolor{amethyst}{rgb}{0.8, 0.0, 0.0}

\definecolor{green}{rgb}{0.0, 0.8, 0.0}

\newcommand{\kms}{km s$^{-1}$}

\newcommand{\rev}[1]{{#1}}

\def\mum{{\mu \rm{m}}}
\def\OI{{\rm [OI]}\, 63 \mum}
\def\OIdue{{\rm [OI]}\, 145 \mum}
\def\CII{{\rm [CII]}\, 158 \mum}

\def\cm-3{{\rm cm^{-3}}}

\def\coldsim{\textsc{ColdSIM} }

\begin{document} 

   \title{High-$z$ [OI] emission lines: \\ \coldsim simulations and ALMA observations}


    \author{M.~Parente \inst{1,2}\orcidlink{0000-0002-9729-3721} 
    \and M.~Bischetti \inst{3,2} \orcidlink{0000-0002-4314-021X}
    \and U.~Maio \inst{1,2} \orcidlink{0000-0002-0039-3102}
    \and F.~Salvestrini \inst{1,2} \orcidlink{0000-0003-4751-7421}
    \and C.~Feruglio \inst{1,2}\orcidlink{0000-0002-4227-6035}
    \and G.~L. Granato \inst{1,2,4}\orcidlink{0000-0002-4480-6909}
    \and C.~Ragone-Figueroa \inst{4, 1, 2}\orcidlink{0000-0003-2826-4799}
    \and R.~Tripodi \inst{5, 6, 2}\orcidlink{0000-0002-9909-3491}
    \and C.~De Breuck\inst{7}\orcidlink{0000-0002-6637-3315}
    \and C.~Ferkinhoff \inst{8}\orcidlink{0000-0001-6266-0213}
    \and L.~Tornatore\inst{1, 9}\orcidlink{0000-0003-1751-0130}}

    \institute
   {INAF, Osservatorio Astronomico di Trieste, via Tiepolo 11, I-34131, Trieste, Italy
    \and
    IFPU, Institute for Fundamental Physics of the Universe, Via Beirut 2, 34014 Trieste, Italy
    \and 
    Dipartimento di Fisica, Università di Trieste, Sezione di Astronomia, Via G.B. Tiepolo 11, I-34131 Trieste, Italy
    \and
    IATE - Instituto de Astronom\'ia Te\'orica y Experimental, Consejo Nacional de Investigaciones Cient\'ificas y T\'ecnicas de la\\ Rep\'ublica Argentina (CONICET), Universidad Nacional de C\'ordoba, Laprida 854, X5000BGR, C\'ordoba, Argentina
    \and
    INAF - Osservatorio Astronomico di Roma, Via Frascati 33, I-00078 Monte Porzio Catone, Italy
    \and
    University of Ljubljana FMF, Jadranska 19, 1000 Ljubljana, Slovenia
    \and
    European Southern Observatory, Karl-Schwarzschild-Strasse 2, 85748 Garching, Germany
    \and
    Department of Physics, Winona State University, Winona, MN 55987, USA
    \and 
    ICSC - Italian Research Center on High Performance Computing, Big Data and Quantum Computing, via Magnanelli 2, 40033,
Casalecchio di Reno, Italy
    \\
    \\
    \email{massimiliano.parente@inaf.it -- mparente270@gmail.com}}

   \date{Received XXX; accepted YYY}

 
  \abstract
   {Neutral-oxygen [OI] far-infrared emission lines at $63 \, \mum$ and $145 \, \mum$ are powerful probes of the physical conditions in the interstellar medium, although they have not been fully exploited in high-redshift studies.}
   {We investigate the connection between [OI] emission lines and key galaxy properties, such as star formation rate (SFR) and H$_2$ content. Our predictions are compared with existing observations and new data analysed in this work.}
   {We post-process the outputs of the \coldsim cosmological simulations with the DESPOTIC model, taking into account $\OI$ self-absorption by cold foreground material. A Random Forest algorithm is employed to accelerate computations and new observational ALMA data for galaxies at redshift $z\simeq 5$-7 are used to validate our model.
   }
   {Our predictions show significant $\OI$ luminosities ($\approx 10^8\,\rm L_\odot$) for galaxies with SFRs of $\approx 10^2\,\rm M_\odot\,{\rm yr}^{-1}$. The $\OIdue$ line \rev{ luminosity is typically $15 \%$ the $\OI$ one
   } and is a factor $\approx 2-20$ below high-$z$ observations. Both [OI] lines correlate with SFR and molecular mass, but exhibit flattening in scaling relations with metallicity and stellar mass. Foreground self-absorption reduces the $\OI$ flux by a factor of $2-4$, consistent with empirical corrections in observational studies. We find typical line ratios of $\OI / \CII \approx 1$ and $\OIdue / \CII \approx 0.2$ -- consistent with $z\gtrsim 6$ observations, but only when $\OI$ self-absorption is included. 
   }  
   {Both $\OI$ and $\OIdue$ lines serve as tracers of star formation and molecular gas at high redshift. Their joint detection can provide constraints on the properties of the early interstellar medium and self-absorption of the $\OI$ line.}


   \maketitle
%

\section{Introduction}

Star formation in galaxies occurs within the densest regions of the interstellar medium (ISM), specifically in molecular clouds (MCs), where efficient cooling is crucial. This cooling is mainly driven by metals present in various forms: solid (e.g., dust), molecular (e.g., CO), and atomic (e.g., C, O, N, Si). Both molecules and atoms are responsible for far-infrared (FIR) fine-structure lines originating from photo-dominated regions (PDRs), which are among the most important coolants \citep[e.g.,][]{Tielens85, Wolfire22}. Thanks to advanced observational facilities that have been active over the past decade, such as ALMA and NOEMA, these FIR lines can be detected and used to probe the physical properties of both local and high-redshift ($z$) galaxies \citep[e.g.][]{Peng25a, Peng25b}.
\\
Among metals, oxygen is the most abundant in the Universe and one of the earliest produced in significant quantities, primarily via core-collapse supernovae (SNII). For this reason, oxygen is often employed as a reliable tracer of global metallicity in galaxies.
In the context of photo-dominated regions (PDRs), oxygen plays a crucial role as a coolant for both cold ($\lesssim 10^4\,\rm K$) and warm ($\sim$10$^4$–10$^6\,\rm K$) gas. In the former case, due to its ionization potential of $13.62\,\rm eV$, which is nearly identical to that of hydrogen, atomic oxygen is confined to neutral regions and exhibits two key FIR fine-structure transitions. The brighter and more prominent line is the $^3P_1 - ^3P_2$ transition at $63.2\,\mu\rm m$, with a critical density of $\approx 5 \times 10^5\,\rm cm^{-3}$ at $T \approx 300\,\rm K$ and an upper level energy of $E_u/k \approx 228\,\rm K$. This transition is a dominant cooling channel in FIR \citep{Malhotra01} and is often optically thick or absorbed by cold foreground material, as revealed by early PDRs observations \citep{Stacey83, Poglitsch96}.
The second line, associated with the $^3P_0 - ^3P_1$ level transition, lies at $145.5,\mu\rm m$ and has an upper-level energy of $E_u/k \approx 327\,\rm K$. This line is generally weaker than the $63\,\mum$ line; thus, simultaneous observations of both lines provide important diagnostics of the gas temperature and the magnitude of $\OI$ self-absorption.
\\
Both $\OI$ and $\OIdue$ lines have been detected in extragalactic observations, thanks to the Photodetector Array Camera and Spectrometer (PACS) onboard \textit{Herschel}, and more recently with ALMA. \cite{DL14} studied the reliability of the $\OI$ line as a tracer of the star formation rate (SFR) in local galaxies, finding it to perform well -- often better than the widely used $\CII$ line in sub-solar metallicity systems. \cite{DiazSantos17} performed a systematic study of FIR emission lines in local luminous infrared galaxies (LIRGs), remarking the presence of self-absorption based on the low inferred $\OI/\CII$ ratios in some sources.
At high redshift, observations of [OI] lines are far more limited and typically focus on lensed, bright submillimeter galaxies (SMGs) or luminous quasars \citep[][]{Sturm10, Coppin12, Brisbin15, Wardlow17, Zhang18, DeBreuck19, Wagg20, Litke22, FA24, Herrera25, Decarli25}. These studies use [OI] lines alongside $\CII$ to better constrain the physical conditions of the high-$z$ ISM and provide insight on $\OI$ optical thickness when both lines are detected.
Notably, recent detections of $\OIdue$ in a small sample of $z \approx 7$ main-sequence galaxies \citep{Fudamoto25} and the most distant $\OI$ detection in the $z=6.04$ quasar J2054-0005 \citep{Ishii2025} have been reported. In the latter case, the authors also suggested evidence of self-absorption based on their line profile analysis.\\
On the theoretical side, several efforts have been made to model FIR emission lines within the framework of galaxy evolution simulations, both semi-analytic \citep[e.g.,][]{Lagos12, Lagache18, Popping19, Yang22} and hydrodynamic \citep[e.g.,][]{Hernandez17, Olsen17, Pallottini17, Katz2019, Lupi20, RP23, Khatri2024, Garcia2024}. This is a challenging task, as the resolution of such simulations is typically too coarse to resolve the individual molecular clouds where PDRs develop. A common strategy is to post-process simulation outputs using dedicated pipelines \citep{Olsen17, Khatri24mod} or specialized PDRs codes -- such as CLOUDY \citep{Ferland17}, UCL–PDR \citep{UDR_code}, DESPOTIC \citep{Krumholz2014} -- which model the chemical and thermal states of sub-grid clouds. As is common with sub-resolution modeling, this process requires assumptions to connect the large-scale simulation outputs with the conditions inside unresolved structures, introducing a degree of arbitrariness that has been explored in various studies \citep[e.g.,][]{Popping19, Lupi20}.
\\
Neutral oxygen lines, however, have received limited attention in these works, with notable exceptions being \cite{Olsen17, Lupi20, Pallottini22, RP23}, of which only the latter is based on cosmological volumes. Motivated by this gap and by the growing observational interest in high-redshift neutral oxygen emission, we present predictions for the $\OI$ and $\OIdue$ lines at $z \geq 6$, based on the \coldsim state-of-the-art cosmological simulations \citep{Maio22} coupled with the DESPOTIC code \citep{Krumholz2014}. Importantly, we introduce for the first time in this context a simplified model to account for $\OI$ self-absorption by cold foreground gas. We compare our predictions to the few existing observations as well as new $z \geq 4.7$ data analyzed in this work and highlight the importance of simultaneously detecting both lines in constraining self-absorption and estimating molecular gas masses at high redshift.\\
The paper is organized as follows. In Sect. \ref{sec:methods} we describe the \coldsim simulations used, along with the pipeline developed to model line emission and $\OI$ self-absorption. Sect. \ref{sec:obs} presents the observational data used for comparison, with particular focus on the new analysis conducted for four objects with $\OIdue$ detections. Our main results -- namely, predictions for self-absorption, 
luminosity functions, scaling relations and line ratios -- are presented in Sect. \ref{sec:results} and interpreted in light of the
existing literature in Sect. \ref{sec:discussion}. Finally, we summarize our findings and conclude in Sect. \ref{sec:conclusions}.

\section{Methods}
\label{sec:methods}
In this section, we describe the \coldsim hydrodynamic chemistry simulations, along with the DESPOTIC code, the machine learning-based pipeline adopted to speed up computations, and the $\OI$ self-absorption model.

\subsection{\coldsim cosmological simulations}

The \coldsim cosmological simulations are a set of numerical simulations focused on the study of the evolution of primordial galaxies, and are extensively presented and discussed in \cite{Maio22}. These simulations are based on $N-$body and hydrodynamical calculations carried out with an extended version of the \textsc{P-Gadget3} code, a non-public evolution of \textsc{Gadget} \citep[][]{GADGET05}. The main feature of these simulations is the inclusion of a detailed time-dependent, non-equilibrium atomic and molecular chemical network to model ionization, dissociation, and recombination processes \citep{Abel97, Yoshida03, Maio07} for the following species: e$^-$, H, H$^+$, H$^-$, He, He$^+$, H$^{2+}$, H$_2$, H$_2 ^+$, D, D$^+$, HD, and HeH$^+$. This network includes the formation of H$_2$ molecules in both primordial and metal-enriched environments, in the latter case catalyzed by dust.
\\
Star formation is modeled stochastically as follows from \cite{Maio09, Maio11}.
This is a modification of the \cite{SH03} model that generates collisionless stellar particles with \cite{Salpeter55} IMF for both population~III and population~II-I regimes.
The star formation rate depends on the H$_2$ cooling and density within each star-forming gas particle.
This latter is responsible for the gas enrichment of individual heavy elements --  C, N, O, Ne, Mg, Si, S, Ca, and Fe -- by Asymptotic Giant Branch stars, Type II and Type Ia SNe, along with their spreading throughout gas particles \citep{Tornatore07, Maio10}. Fine-structure lines from the most relevant metals are also included in the gas cooling budget \citep{Maio07}.
A number of other relevant processes for H$_2$ chemistry and cooling, such as HI or H$_2$ self-shielding, UV background, photo-electric and cosmic-ray (CR) heating, are also taken into account (see \citealt{Maio22} for full details).
\\
The simulation box used in this work is the largest among the \coldsim runs, enabling predictions for the most massive objects comparable to those observed at high redshift. 
A standard $\Lambda$CDM cosmology is assumed, with a present-day Hubble parameter normalized to 100~km/s/Mpc $h=0.7$ and baryon, matter and $\Lambda$ density parameters $\Omega_{0,b} = 0.045$, $\Omega_{0,m} = 0.274$, and $\Omega_{0,\Lambda} = 0.726$, respectively.
The box has a comoving size of $ 50\,{\rm Mpc} / h $ initialized with $ 2 \times 1000^3$ particles. Dark-matter and gas particle masses are
$ 1.1 \times 10^7 \, M_\odot$ and $ 2.3 \times 10^6 \, M_\odot$, respectively.
Gravitationally bound substructures are identified by means of friends-of-friends
and substructure-finder
algorithm \citep{Dolag09}.
We select galaxies with at least $300$ members (stars$+$gas), whose properties are computed considering gas and stellar bound particles in $0.1 R_{200}$, with $ R_{200}$ virial radius.

\subsection{Modeling lines emission}
\label{sec:mod:lines}

From each gas particle in our simulated galaxies, we consider five quantities to predict lines luminosities: mass $M_{\rm part}$, radius\footnote{Derived assuming particles are spheres of volume  
$V_{\rm part} = M_{\rm part}/\rho_{\rm part}$, with $\rho_{\rm part} $ simulated particle density.} $R_{\rm part}$, star formation rate SFR$_{\rm part}$, oxygen [O/H] $_{\rm part}$ and carbon [C/H]$_{\rm part}$ particle abundance. These quantities are used to set up the DESPOTIC post-processing detailed in the following.
We note that we only consider gas particles that host star formation and we need a post-processing approach to account for unresolved molecular clouds.
We neglect diffuse, hot particles, since we have checked that their contribution is negligible due to their typically low densities ($n \lesssim 1 \, \cm-3 $) and high temperatures ($T\gtrsim 10^{4}\, {\rm K}$, where most of oxygen is ionized).

\subsubsection{DESPOTIC set-up and luminosity computation}

Line luminosities are obtained by coupling the output of our simulation with DESPOTIC \citep{Krumholz2014}. This code simultaneously computes the statistical levels, thermal, and chemical equilibrium of clouds, provided with the ingredients and assumptions described below. For a detailed description of the model, we refer the reader to the main paper, as well as to \cite{Narayanan2017} and \cite{Garcia2024} -- from which we adopt parameters not explicitly specified here (see their Table 1). In this section, we focus only on the main features and the adaptation of the simulation data for use as DESPOTIC inputs.
\\
Clouds (i.e., particles) are treated as spheres of uniform density $n_{\rm H}$ divided into $N_{\rm zones}=40$ concentric shells \footnote{We have checked that this number is sufficient to reach convergence.}, each of which hence with different column densities. The densities are derived from the simulation and are in the range $n_{\rm H}\approx 1-10^4 \, {\rm cm}^{-3}$.
We provide only the total mass of the clouds, without making any assumptions about the fraction of mass in neutral or molecular form, as this is directly computed by DESPOTIC.
\\
These clouds also include dust, which is essential for the energetic balance of the cloud. Our particles are assigned a dust-to-metals (DTM) ratio from their metallicity $Z_g$, exploiting the highest-redshift fit for DTM vs $Z_g$ suggested by \cite[][$4 < z < 5.3$]{Popping2022}. The dust parameters, i.e. gas-dust coupling coefficients and dust-radiation cross sections are assumed to scale with DTM (normalized to the Milky Way value, DTM$_{\rm MW}=0.44$) and are the same as those adopted in \cite{Garcia2024} (see their Tab. 1).
The chemical composition of each cloud is computed exploiting the Gong-Ostriker-Wolfire (GOW, \citealt{Gong2017}; see also \citealt{Glover2012}) network, a descendant of the \cite{Nelson1999} network for C-O chemistry and the \cite{Glover2007} network for hydrogen. The oxygen and carbon abundances given as input to the network are those extracted from the simulation, which follows elemental abundance for each particle. 
More details can be found in \cite{Narayanan2017} and references therein.
\\
Two crucial ingredients are derived directly from the simulation: the modeling of the interstellar radiation field $\chi_{\rm FUV}$ (ISFR) for photo-reactions and the cosmic-ray rate $\zeta_{\rm CR}$.
The FUV ISRF is normalized to the MW value and scales with the particle SFR density $\Sigma_{\rm SFR}$ attenuated by dust via an escape fraction $\beta_{\rm dust}$:
\begin{equation}
   \frac{\chi_{\rm FUV}}{\chi_{\rm FUV, \, MW}}  = \Sigma_{\rm SFR} \cdot \frac{\beta_{\rm dust}}{\beta_{\rm dust, \, MW}}.
\end{equation}
The dust-dependent escape fraction is computed for each particle as $\beta_{\rm dust} = (1-e^{-\tau_{\rm FUV}})  / \tau_{\rm FUV}$, where
\begin{equation}
    \tau_{\rm FUV} = \kappa_{\rm abs}({\rm DTG}) \cdot \frac{3 M_{\rm dust}}{4 R^2_{\rm part}}.
\end{equation}
In the previous equation $\kappa_{\rm abs}({\rm DTG})$ is the dust absorption coefficient in the FUV set to $1.078 \cdot 10^5 \, {\rm cm}^2 {\rm g}^{-1}$ \citep{Draine11} and assumed to scale linearly with the DTG \citep{Ferrara22}, and $M_{\rm dust}$ is the dust mass of each particle. As for MW values, we adopt $\Sigma_{\rm SFR, \, MW}=0.79 \cdot 10^{-3}\, M_\odot {\rm yr}^{-1} \, {\rm pc^{-2}}$ and $\beta_{\rm dust, \, MW} \sim 0.5$ \citep{Bonatto2011, Garcia2024}.
\\
As for the cosmic-ray (CR) rate, an admittedly poorly-known parameter even in the local neighborhood, we follow \cite{Krumholz2023}. They link the SFR-induced CR rate to the depletion time $t_{\rm dep}$ via 
$ \zeta_{\rm CR} \simeq 10^{-16} \, {\rm s^{-1}} / (t_{\rm dep}[{\rm Gyr}]). $
In our modeling, the depletion time is the ratio between gas particle mass and SFR, $t_{\rm dep} = M_{\rm gas}/{\rm SFR}$.
This parametrization is quite different from another common technique adopted to model CRs, that is scaling the rate with the \textit{unshielded} SFR density.
As the aforementioned authors noted, while the $\Sigma_{\rm SFR}$ can differ by various orders of magnitude across the galaxy population (e.g. $\Sigma^{\rm starburst}_{\rm SFR} \sim 10^5 \Sigma^{\rm MW}_{\rm SFR}$), depletion times typically do not.
We get back to this point in Sect. \ref{sec:modvar}.
\\
%
%
Once all cloud properties are defined, DESPOTIC solves for chemical, thermal, and statistical equilibrium simultaneously in each zone of the cloud, treating the zones independently. 
The chemical abundances are evolved over time using reaction rates from the adopted network, while thermal equilibrium is achieved by balancing heating and cooling processes for both gas and dust.
The level populations of atomic and molecular species are determined by equating the rates of population and de-population due to collisions and radiative processes. Radiative trapping is accounted for using the escape probability formalism, and collisional terms are adjusted for clumping based on non-thermal velocity dispersion.
\\
Once convergence is reached in all zones, DESPOTIC provides the equilibrium gas and dust temperatures, energy balance, chemical abundances, and level populations. Line luminosities are then computed by summing the contributions from each transition across all cloud zones, including CMB corrections.

\subsubsection{Extrapolating results with Machine Learning}

\label{sec:ML}

We run DESPOTIC on a representative subset of gas particles ($\approx 10^5$), randomly selected from galaxies spanning the full range of stellar masses. The outputs from these simulations are then used to train a Random Forest Regressor, which is subsequently applied to predict DESPOTIC-like results for the remaining gas particles in the selected galaxies. This approach significantly reduces the computational cost associated with running DESPOTIC on the full particle set, while maintaining a high level of accuracy, as demonstrated below. Before describing the pipeline in detail, we note that similar strategies are increasingly common in studies of FIR emission lines, as seen in \cite{Katz2019}, which uses CLOUDY, and \cite{Garcia2024}, which employs DESPOTIC.
\\
In this work, we train multiple Random Forest (RF) models to predict line luminosities, chemical abundances, and gas temperatures from simulation data, using five
input features: gas particle mass, radius, carbon and oxygen abundances, and SFR density. The labeled dataset (i.e., particles with known luminosity, temperature, and chemical composition) is randomly divided into a training set ($80 \%$) and a test set ($20 \%$). We use the \texttt{RandomForestRegressor}\footnote{\url{https://scikit-learn.org/stable/modules/generated/sklearn.ensemble.RandomForestRegressor.html}{}} from the \texttt{scikit-learn} Python library, with $200$ estimators (trees), and no maximum tree depth constraint.
The trained RFs consistently achieve high $R^2$ scores ($\gtrsim 98-99 \%$), indicating excellent agreement between predicted and true values, with $100 \%$ corresponding to a perfect match. We briefly expand on the accuracy of this pipeline in App. \ref{app:RF}.


\subsubsection{Modeling foreground $\OI$ self-absorption}

\label{sec:mod:sa}

Although the $\OI$ is among the strongest lines in typical ISM conditions, it is also optically thick and thus susceptible to self-absorption, which can reduce its observed intensity. This reduction arises from the presence of cold, sub-thermally excited material along the line of sight -- either associated with the source itself or with intervening foreground clouds \citep[e.g.][]{Goldsmith2019}. The net effect of this self-absorption is an increase in the $\OIdue/\OI$ line ratio, often $\gtrsim 0.1$. Such elevated ratios are only explainable within a scenario where the $\OI$ emission is optically thick, while the $\OIdue$ line is practically always thin. Such an increased ratio has been observed both in velocity-resolved studies of star-forming regions and in integrated emission from external galaxies \citep[e.g.][]{Caux1999, Leurini2015, Farrah13, FernandezOntiveros2016}.
The common strategy adopted in observational studies to account for self-absorption is to apply a correction factor of $2-4$ to the observed $\OI$ intensity \citep[e.g.][]{Schneider2018, Goldsmith2021, Ishii2025}. In this work, we attempt a more physically motivated, albeit simplified, treatment of foreground self-absorption by exploiting predictions from cosmological simulations regarding the spatial distribution of gas.
\\
For each emitting gas particle, we estimate self-absorption along the line of sight by accounting for all the gas particles located within a cylindrical volume with radius equal to the particle $R_{\rm part}$.
Each gas particle is characterized by a center-of-mass velocity $\textbf{v}_{\rm part}$ and a one-dimensional velocity dispersion computed as \citep{Olsen2015, Khatri2024}:
\begin{equation}
    \sigma_v [{\rm km \, s^{-1}}] = 1.2 \left( \frac{M_{\rm part}}{290 \, M_\odot} \right)^{1/2} \left( \frac{R_{\rm part}}{\rm pc} \right)^{-1/2}.
\end{equation}
Foreground particles contribute to the absorption of the $\OI$ line emitted by a given particle with velocity $\textbf{v}_{\rm part, em}$ and dispersion ${\sigma_v}_{\rm part, em}$ if their velocity profiles overlap:
%
\begin{equation}
    \left| v^{\hat{z}}_{\rm part, em} -v^{\hat{z}}_{\rm part} \right| \leq 2 \sqrt{{\sigma_v}^2_{\rm part, em} + {\sigma_v}^2_{\rm part}}.
    \label{eq:velcond}
\end{equation}
All particles satisfying this criterion contribute to the total neutral oxygen column density, $N_{\rm OI}$, used to compute the foreground optical depth \citep{Goldsmith2019, Goldsmith2021}:
\begin{equation}
    \tau_{\rm fore} = \frac{N_{\rm OI} [{\rm cm}^{-2}]}{2.03 \cdot 10^{17} \Delta v _{\rm abs} [{\rm km \, s^{-1}}]},
\end{equation}
where $\Delta v_{\rm abs} = \sqrt{8 {\rm ln}(2)} {\sigma_v}$ is the FWHM of the line. The line intensity is then attenuated by a factor of $(1-e^{-\tau_{\rm fore}})/\tau_{\rm fore}$.\\
The $\CII$ line can also exhibit self-absorption \citep{Gerin15}, although this effect is much less significant than for the $\OI$ line; therefore, we neglect it in this work.

\section{Observational dataset}

\label{sec:obs}

\begin{table*}[]
    \setlength{\tabcolsep}{3pt}
    
    \caption{\small Physical properties of the newly analysed data. All measurements have been corrected for magnification where needed. References are provided for the magnification factor, as well as for the $\OI$ and $\CII$ luminosities. References to the SFRs correspond to the IR luminosities used for their derivation. Molecular gas masses are inferred from the $\CII$ luminosity (see text for details).
    }
    \centering
    \begin{tabular}{lccccccc}    
    \hline
    ID & $z$ & ${\rm log } \, (L_{\OIdue}/{\rm L_\odot}) $ & ${\rm log } \, (L_{\OI}/{\rm L_\odot})$ & ${\rm log } \, (L_{\CII}/{\rm L_\odot})$  & ${\rm log} \, (M_{\rm H_2}/{\rm M_\odot})$ & ${\rm SFR} \, [{\rm M_\odot / {\rm yr}}]$ & $\mu$\\
    \hline
    A1689$-$zD1 & $7.132$ & $8.28 \pm 0.06$ & $-$ & $9.04 \pm 0.04^a$ &  $10.5 \pm 0.3$ & $84 \pm 3 ^b$ & $4.14 \pm 0.36 ^c$ \\
    SPT0346$-$52 & $5.656$ &  $9.11 \pm 0.43$ & $<9.72 ^d$ & $9.53 \pm 0.06^d$ & $11.0 \pm 0.3$ & $6228 \pm 519^e$ & $ 5.6 \pm 0.1^f$ \\
    J204314$-$214439 & $4.680$ &  $8.91 \pm 0.12$ & $-$ & $9.77 \pm 0.02$ & $11.1 \pm 0.3$ & $1674 \pm 138$ & $ \sim 3^g$ \\
    J2054$-$0005 & $6.038$ &  $8.95 \pm 0.01$ & $9.65 \pm 0.14^h$ & $9.53 \pm 0.06^i$ & $10.1 \pm 0.1$ & $950 \pm 221 ^l$ & $ - $ \\

    \hline
    \label{tab:physpropobs}
    \end{tabular}  
        \caption*{\small 
        References: \textit{(a)} \cite{Killi23} \textit{(b)} \cite{Akins22} \textit{(c)} \cite{Fudamoto25} \textit{(d)} \cite{Litke22} \textit{(e)} \cite{Ma2015} \textit{(g)} \cite{Zavala2015} \textit{(h)} \cite{Ishii2025}  \textit{(i)} \cite{Wang2013} \textit{(l)} \cite{Salak24}}
        \vspace{-0.5cm}
\end{table*}

We compare our results with newly analyzed observational data and existing literature, as detailed below.
Unless stated otherwise, SFRs are inferred from IR luminosities using the conversion coefficient $1.73 \times 10^{-10} \, \rm M_\odot \, {\rm yr^{-1}} / L_\odot$ \citep{Kennicutt12}, which assumes a Salpeter initial mass function (IMF). For the existing literature, we apply a scaling factor of 1.6 to values originally derived using a Chabrier or Kroupa IMF. In the case of lensed sources, we correct the IR and line luminosities using a single magnification factor $\mu$ per galaxy (Table \ref{tab:physpropobs}). \rev{The decision not to adopt differential lensing is motivated by its generally limited impact \citep[e.g.][]{Apostolovski19}. For SPT 0346–52 -- one of our targets -- \citet{Litke22} found consistent lens models and magnification factors across multiple line tracers, implying a negligible effect from differential lensing.}
Molecular gas masses ($M_{\rm H_2}$) are inferred from $\CII$ luminosities by adopting the conversion factor by \citep{Zanella18} for star-forming galaxies, and that by \citep{Salvestrini25} for Quasi-Stellar Object (QSO) host galaxies. We consider as uncertainty on $M_{\rm H_2}$ the dispersion reported in these studies, respectively $0.3$ dex and $0.1$ dex.

\subsection{New $z \gtrsim 4.7$ observations}
\label{sec:obs:newdata}

\begin{figure}[]
    \centering \includegraphics[width=0.9\columnwidth]{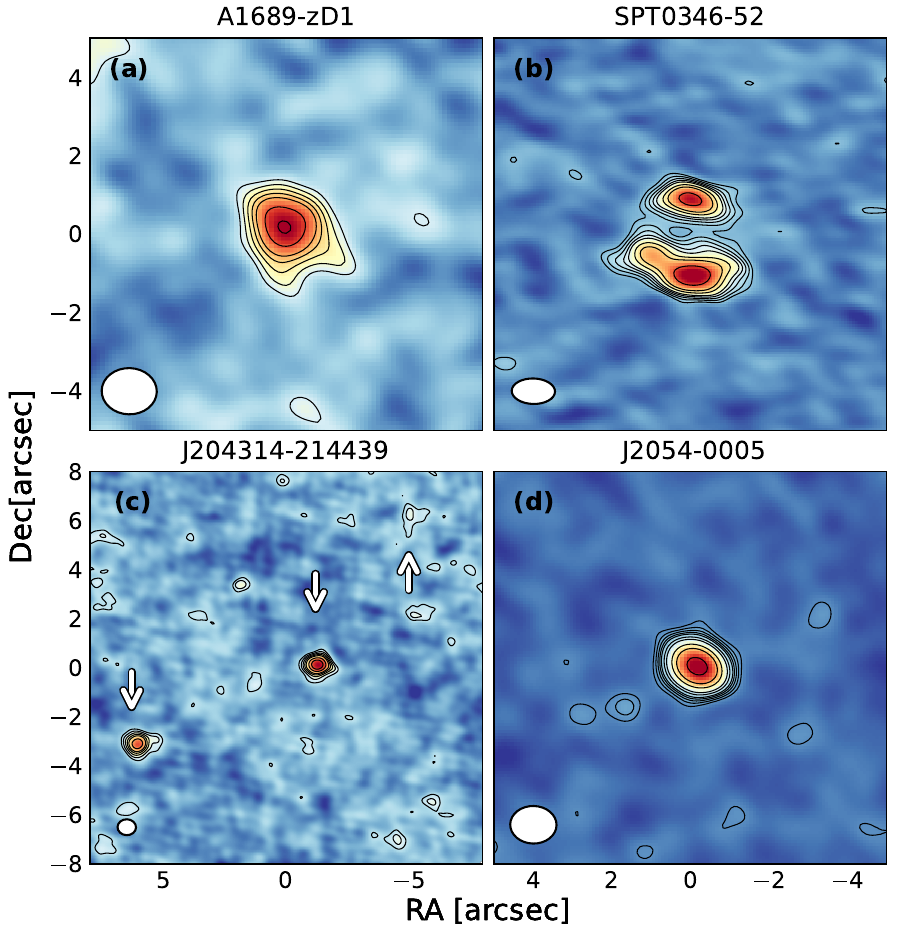}
    \caption{ \small [OI] 145$\mu$m emission maps. Contour lines correspond to [2, 3, 4, 5, 6, 8, 10, 15, 20, 30]$\,\sigma$ confidence levels, where 
    $\sigma = 0.030$, 0.098, 0.082 and 0.024~Jy~beam$^{-1}$~km~s$^{-1}$ for panels a, b, c, and d, respectively. In panel c, arrows indicate the location in which we detect [OI] or [CII] emission at $z=4.680$, in addition to continuum emission (Figure \ref{fig:hlsmaps}).
    In each panel, the bottom-left white ellipse shows the ALMA beam.
    \\
    }
    \label{fig:newoi}
\end{figure}

We include archival data from ALMA project 2023.1.01450.S targeting the $\OIdue$ emission for four galaxies at $z\gtrsim4.7$. This includes three lensed sub-millimeter galaxies, namely A1689$-$zD1 \citep[$z=7.133$,][]{Wong2022}, SPT0346$-$52 \citep[$z=5.656$][]{Litke2019}, and J204314$-$214439 \citep[$z=4.680$,][]{Zavala2015}, and the host-galaxy of quasar J2054$-$0005 \citep[$z=6.038$, ][]{Wang2013}. The analysis of these data is summarized in Appendix~\ref{app:newdata}, and the 0$^{th}$ moment maps of [OI]~145$\mu$m emission are shown in Figure \ref{fig:newoi}. Resulting $\OIdue$ luminosities have been derived from the magnification-corrected fluxes in Table \ref{tab:almadata} using Eq.~(1) in \cite{Solomon2005}. 
\\
We find $\OIdue$ luminosities in the range $2\times10^8-10^9$~$\rm M_\odot$, that are a factor of about 4-7 lower than the [CII] luminosities previously measured for these sources. Although some scatter might be due to the different sensitivities and beam sizes of the [CII] observations listed in Table \ref{tab:physpropobs}, this result suggests that $\OIdue$ typically represents a significant fraction of the total [CII] luminosity.
In J2054$-$0005 the $\OIdue$ luminosity corresponds to about 20\% of that of $\OI$ and, similarly, $\OIdue$/$\OI\gtrsim25 \%$ for SPT0346$-$52.
\\
Data for $\OIdue$ observations of A1689$-$zD1 from ALMA cycle 9 were presented in \cite{Fudamoto25}, with similar angular resolution and about two times lower sensitivity. We measure a consistent $\OIdue$ flux.
Concerning SPT0346$-$52, we detect two bright sources in the [OI] map, the southern of which has an elongated, arc-like shape. We verified that this is consistent with the multiple images detected in high-resolution ALMA band 7 observations by \cite{Litke2019}. 
\cite{Litke22} reported a previous, low-SNR detection of $\OIdue$ in SPT0346$-$52, based on ALMA cycle 2 data with a $\sim0.25$ arcsec angular resolution, and evidence that this galaxy is undergoing a merger. We recover a factor of $\sim4$ higher flux, likely because the low-resolution observation presented in this work better recovers the emission from diffuse gas dispersed by the merger. 
\\
At the ALMA resolution, we detect two distinct $\OIdue$ sources in the map of J204314$-$214439. They show a very similar [OI] line profile centered at the same redshift, which suggests that they are distinct images from the same galaxy, in agreement with the interpretation of this source by \citep{Zavala2015}. 
Previous measurements of $\CII$ and IR luminosities in J204314$-$214439 were based on low resolution observations from SCUBA-2 ($\sim15$ arcsec) and AzTEC/LMT ($\sim8$ arcsec), respectively \citep{Zavala2015}. Accordingly, we created a new 0$^{th}$ moment map of [CII] emission based on ACA data from project 2021.1.00265.S (angular resolution 7.3$\times$5.2 arcsec$^2$, Figure \ref{fig:hlsmaps} left). 
As previous continuum observations did not allow to robustly disentangle the emission of two foreground galaxies along the line of sight from that of J204314$-$214439 \citep{Zavala2015}, we also used the 2023.1.01450.S data to create a high resolution map (0.75$\times$0.65 arcsec$^2$) of the $\sim362.7$ GHz continuum (Figure \ref{fig:hlsmaps} right). We find that the bulk of the continuum emission is located with an elongated source, likely associated with one of the foreground galaxies. Instead, we measure the continuum of J204314$-$214439 by considering only the emission from the multiple images in which we also detect [OI] and [CII] emission at $z=4.680$. 
We derive the IR luminosity of J204314.2$-$214439 by fitting the $\sim362.7$ GHz continuum with a modified black body following the approach in \cite{Tripodi24}, by assuming a dust temperature of $T_{\rm d}=37 \, {\rm K}$ and $\beta=1.70$ \citep{Zavala2015}.
These results are in line with the assumptions made in the simulation run \cite[][]{Maio22}.

\subsection{Existing literature}
We consider the compilation\footnote{Available at 
\url{https://zenodo.org/records/6576202}.} of \cite{RP23} and references therein, including galaxies in a wide redshift range ($0 \leq z \lesssim 6.5$) with the detection of several FIR lines. 
Strong AGN galaxies are not present in this compilation. 
\\
We include the hot dust-obscured galaxy W2246–0526 at $z=4.601$, with both $\OI$ and $\OIdue$ observations from \cite{FA24}, and a SFR estimated from the IR luminosity reported by \cite{Fan18}.
We include the recent $\OI$ detection in the bright QSO J2054-0005 at $z=6.04$ by \cite{Ishii2025}. To date, this is the highest $z$ detection of this line. We also consider the $\OIdue$ detection in the host galaxy of another luminous QSO J2310+1855 at $z=6.003$ by \cite{Li2020}. The upper limit of the $z=5.6$ SPT0346$-$52 $\OI$ luminosity is taken from \cite{Litke22}.
Finally, we include three main sequence star forming galaxies at $6.5 \leq z \leq 7.7$ from the REBELS sample, for which $\OIdue$ were reported by \cite{Fudamoto25}.

\section{Simulation results}
\label{sec:results}

\begin{figure*}
    \centering
    \includegraphics[width=0.45\columnwidth]{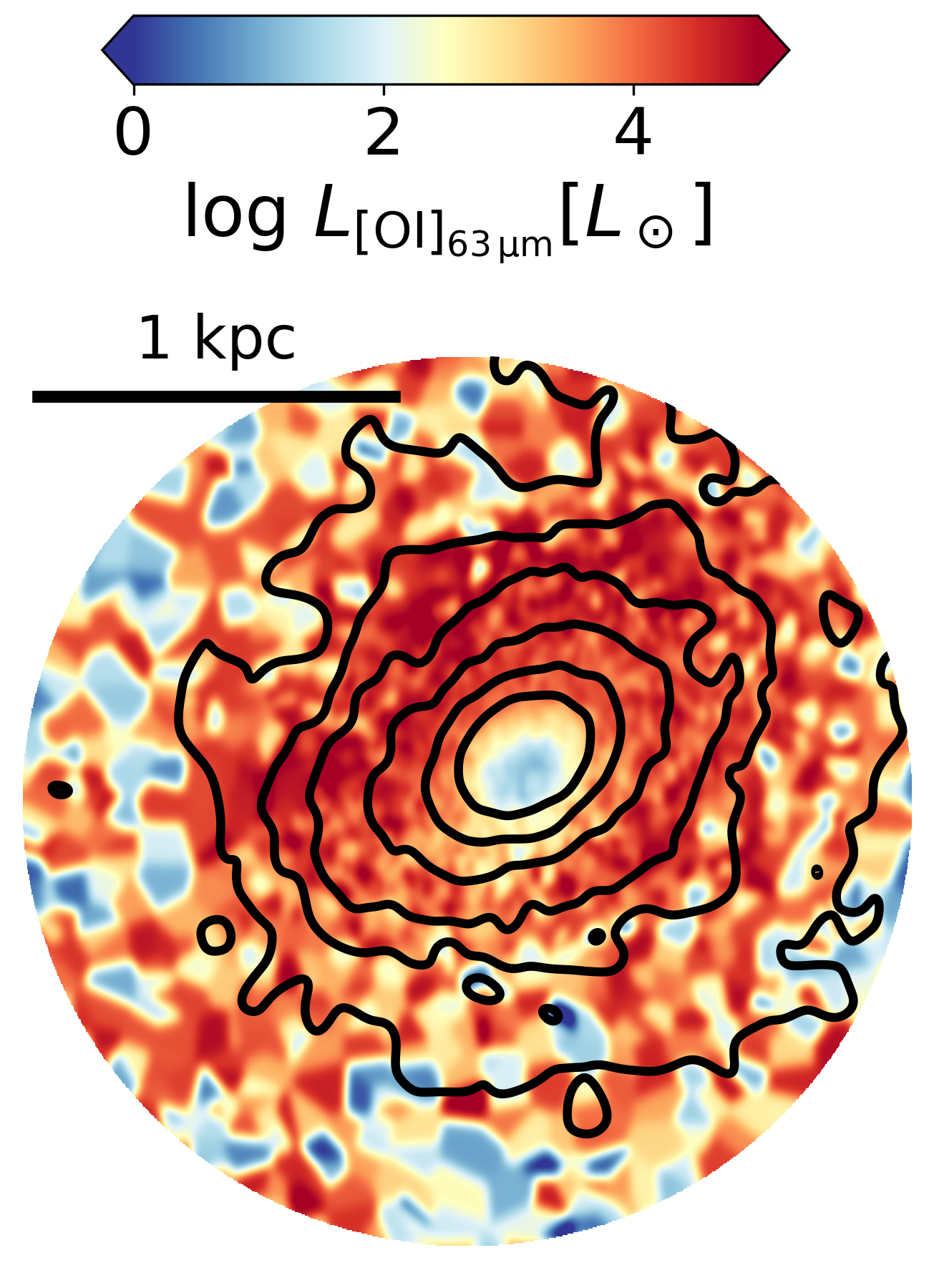}\qquad
    \includegraphics[width=0.45\columnwidth]{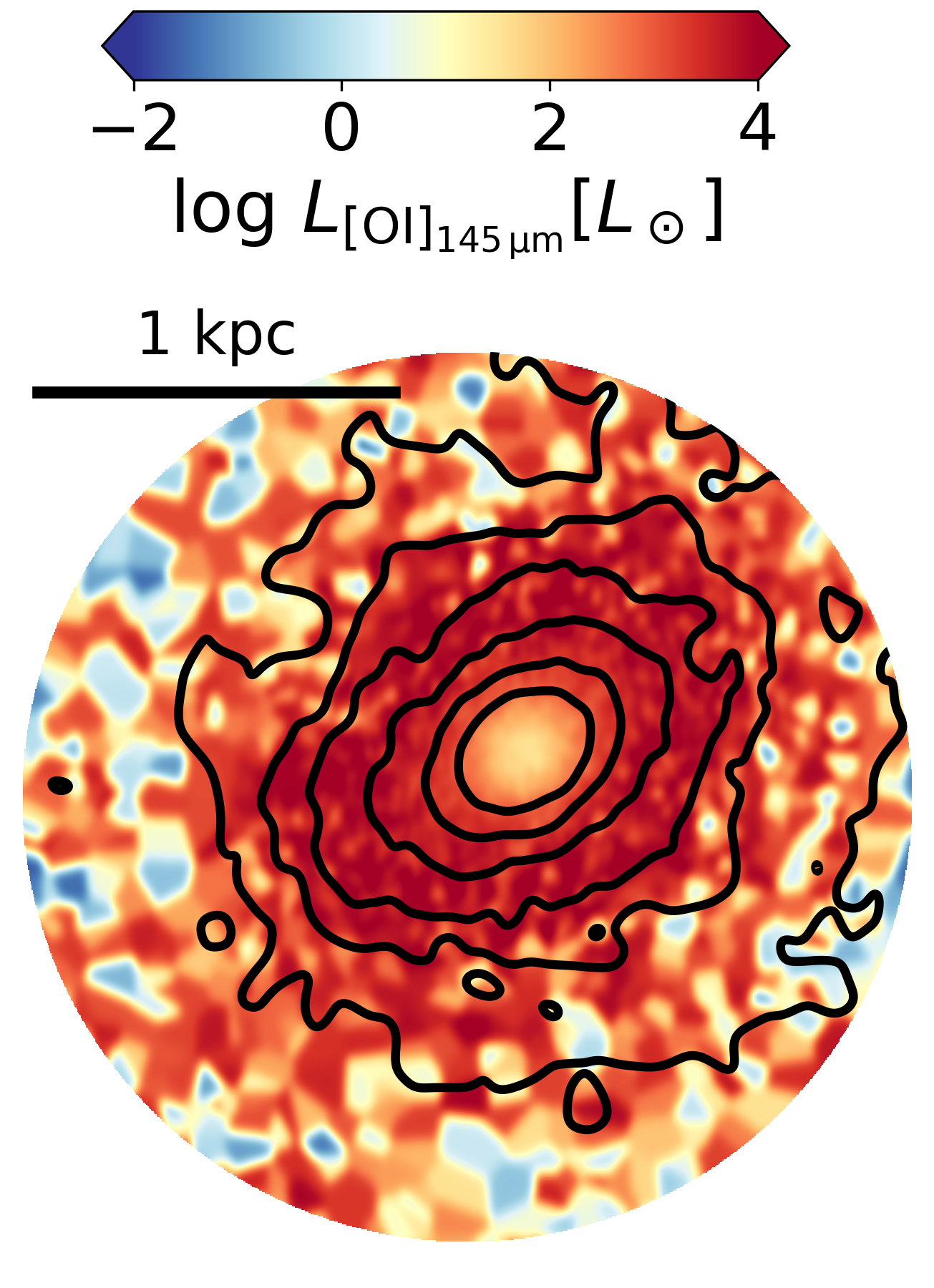}\qquad
    \includegraphics[width=0.45\columnwidth]{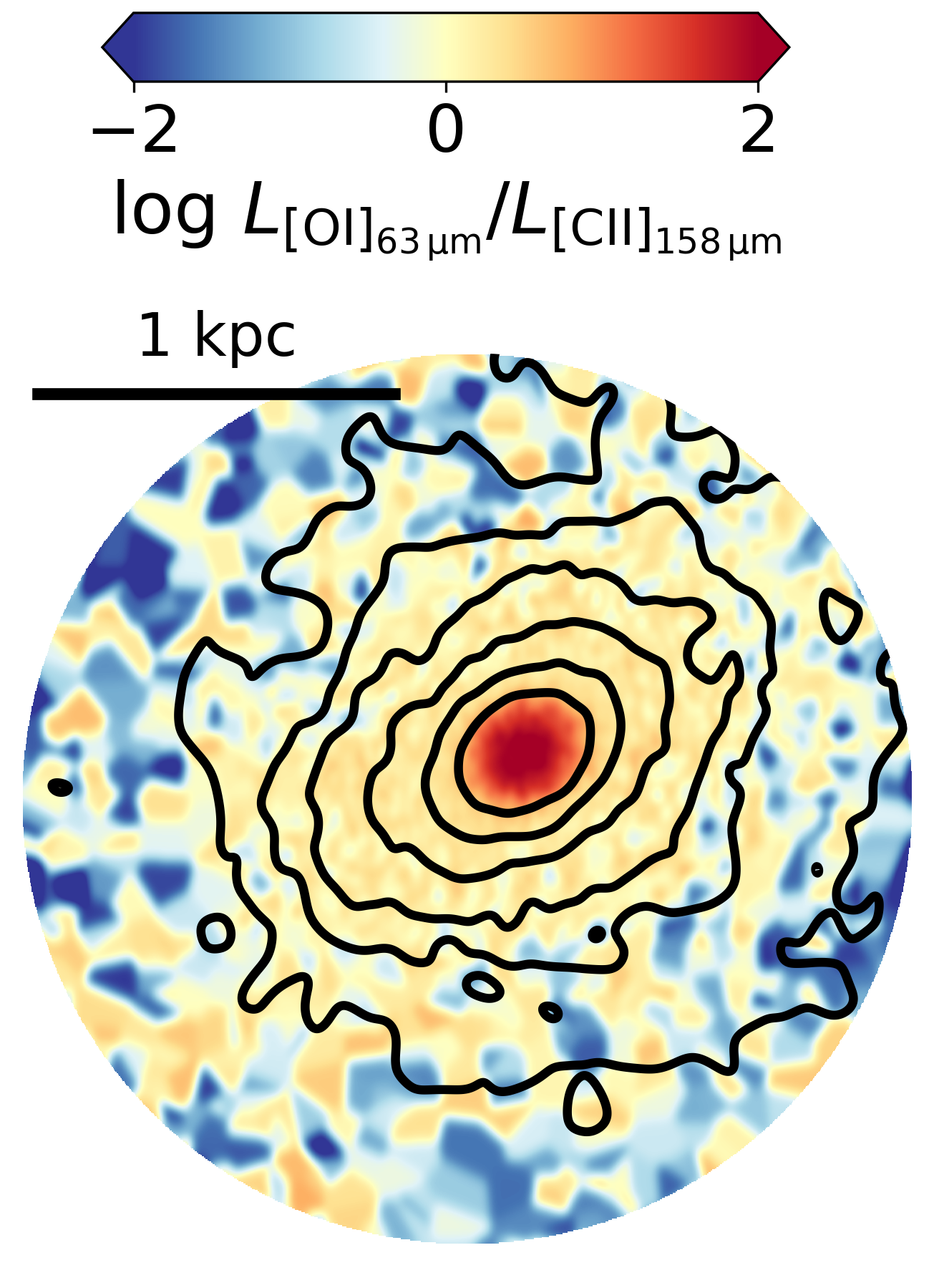}\qquad
    \includegraphics[width=0.45\columnwidth]{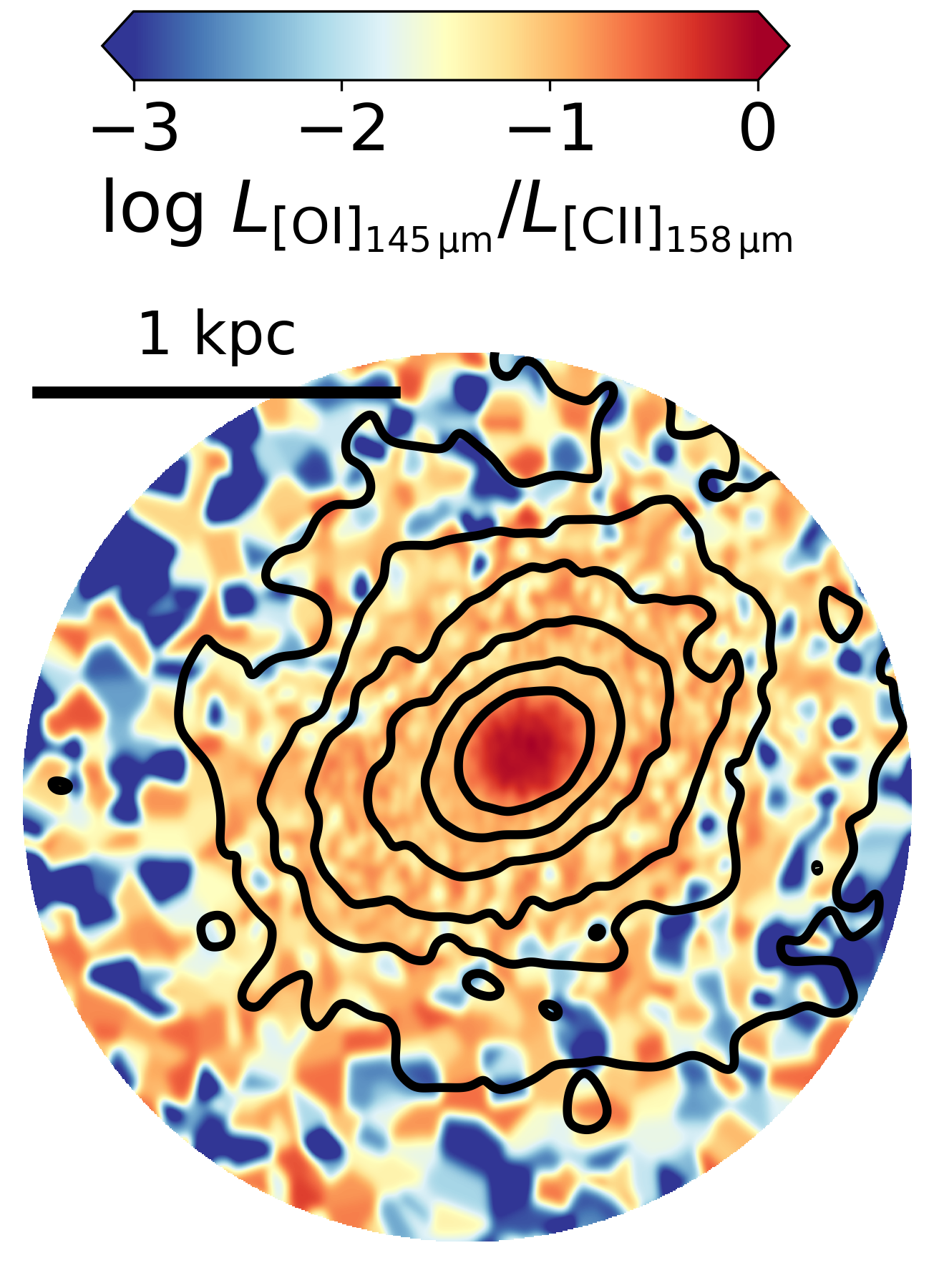}\qquad
    \caption{\small \rev{Visualization of the $\OI$ and $\OIdue$ luminosities, and the $\OI/\CII$ and $\OIdue/\CII$ ratios, for the most massive simulated galaxy at $z=6.14$.} \rev{Black} contours follow the gas density distribution.
    }
    \label{fig:maps}
\end{figure*}

In this section, we present the main results of our work concerning the modeled $\OI$ and $\OIdue$ luminosities of simulated galaxies. We do not present general simulation predictions~-- such as mass or UV luminosity functions, baryon fractions, galaxy main sequence, mass-metallicity relation, etc.~-- that have been discussed in previous works \citep{Maio22, Maioviel23, Casavecchia24, Casavecchia25}.
\\
As an illustrative example of our modeling pipeline, in Fig. \ref{fig:maps} we show the most massive galaxy ($M_{\rm stars}\sim 3.5 \times 10^{10}\,M_\odot$) in our simulation at $z=6.14$, with gas particles color-coded by their $\OI$ and $\OIdue$ luminosities \rev{as well as their $\OI/\CII$ and $\OIdue / \CII$ ratios}. \rev{From the ratios it can be inferred that} $\CII$ emission is extended and diffuse compared to the [OI] lines, which remain faint in the low-density ($n_{\rm H} \lesssim 10^2 \, \mathrm{cm}^{-3}$) outer regions. In contrast, [OI] lines become significantly brighter in dense central regions ($n_{\rm H} \gtrsim 10^4 \, \mathrm{cm}^{-3}$\, $T_{\rm gas} \lesssim 100 \, \mathrm{K}$), where their higher critical densities make them effective tracers of such conditions. 

\subsection{Predictions of $\OI$ self-absorption}

\begin{figure}
    \centering
    \includegraphics[width=0.8\columnwidth]{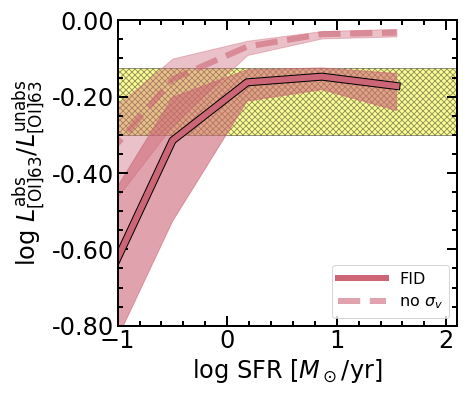}
    \caption{\small Fraction of self-absorbed $63 \, \mum$ [OI] line flux due to foreground material as a function of the galaxy SFR as predicted by our model. Solid lines refer to median trends while the shaded area are the 16-84th percentiles dispersion. The solid line refers to the fiducial model, while the dashed one is obtained without considering the condition on velocity in Eq. \ref{eq:velcond}. The yellow dashed region represent the typical values adopted in observational studies (reduction by a factor of $\sim 2-4$).}
    \label{fig:betaOI}
\end{figure}

The outcome of the simple model for $\OI$ self-absorption introduced in Sect. \ref{sec:mod:sa} is shown in Fig. \ref{fig:betaOI}, where the fraction of $\OI$ flux absorbed by foreground material is shown as a function of the SFR of our model galaxies at $z=6$. We find no significant redshift evolution of this quantity in our model. 
The absorbed fraction increases with SFR, up to $\sim 1 \, \rm M_\odot / yr$ where it flattens and slowly decreases in the largest objects in our volume. The increasing trend of the relation is due to the major [OI] column density found in larger objects. At the same time, the flattening is due to the condition on velocities introduced in Eq. \ref{eq:velcond}. Briefly, the velocity distribution of gas particles broadens faster than the velocity dispersion of GMCs with increasing SFR. As a consequence, at fixed column density, fewer particles in the line of sight can be targeted as absorbers. To prove this, we show in Fig. \ref{fig:betaOI} what would be obtained without considering the Eq. \ref{eq:velcond} condition (no $\sigma_{v}$ case) and in the rest of the paper we will show results also for this version of the model to bracket our predictions.
\\
Interestingly, despite its simplicity, for galaxies with SFR$\gtrsim 0.1 \,\rm M_\odot/$yr our model yields attenuation factors that are broadly consistent with the empirically adopted corrections of $2-4$, as shown in Fig. \ref{fig:betaOI}. Nonetheless, we acknowledge the limitations of this approach. In particular, the absorbed radiation does not contribute to the excitation of the absorbing gas, which is assumed to be cold ($T_{\rm ex} \lesssim 50 \, {\rm K}$), such that only the ground state is significantly populated. A more rigorous treatment would involve solving the full radiative transfer problem iteratively, accounting for the mutual coupling between radiation and level populations in the absorbing material.
However, this would be computationally prohibitive.


\subsection{Luminosity functions}

\begin{figure}
    \centering
    \includegraphics[width=0.8\columnwidth]{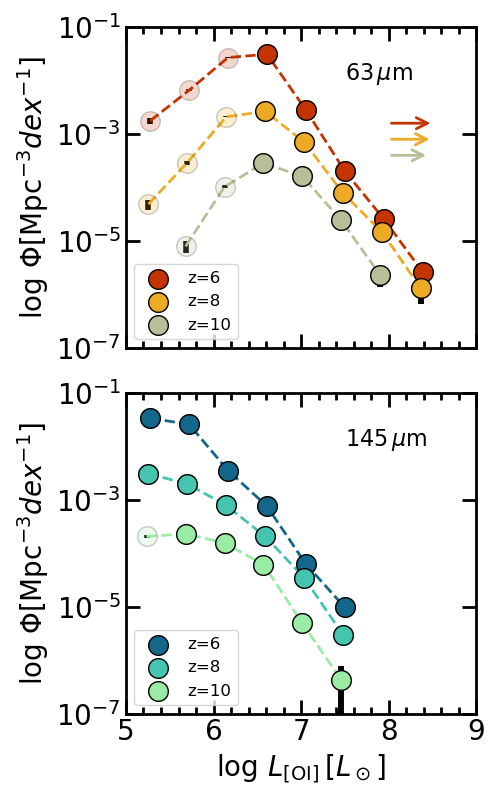}
    \caption{\small Luminosity function of $\OI$ (top panel) and $\OIdue$ (bottom panel) at $z=6,\,8,\,{\rm and} \, 10$ as predicted by our simulation. Errorbars are obtained by bootstrapping. The colored arrows in the top panel indicate the mean magnitude of the foreground self-absorption at each redshift. Low opacity points are those affected by resolution effects.}
    \label{fig:OILF}
\end{figure}

We begin by examining the luminosity functions of $\OI$ and $\OIdue$ as predicted by our simulations at redshifts $z = 6$, $8$, and $10$. These are presented in Fig. \ref{fig:OILF}, excluding data points below the low-luminosity threshold where sample incompleteness becomes relevant. Currently, observational determinations of these luminosity functions at such high redshifts are not available, therefore, our predictions are intended to serve as theoretical benchmarks for future studies that will have sufficient data at $z \geq 6$.
The high-luminosity end of our predictions is well described by a simple power law, exhibiting a non-evolving slope of approximately $-1.9 \pm 0.1$. The impact of $\OI$ self-absorption is very similar at all redshift, with a magnitude $\approx 3$.
The overall normalization increases with cosmic time, keeping track of the ongoing cosmological structure formation.
This suggests an expected redshift evolution for the statistics of both OI lines already at such early epochs.


\subsection{Scaling relations}
\label{sec:scaling}

\begin{figure*}
    \centering
    \includegraphics[width=0.65\columnwidth]{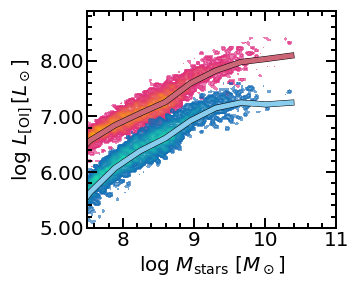}\quad
    \includegraphics[width=0.65\columnwidth]{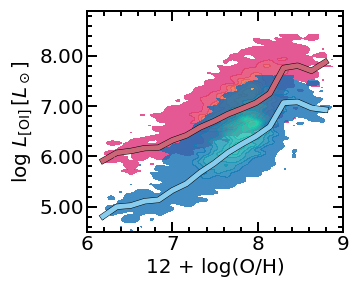}\quad
    \includegraphics[width=0.65\columnwidth]{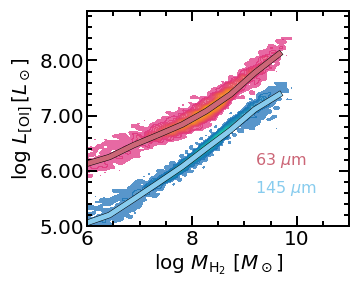}
    \caption{\small Relations between the $\OI$ (red) and $\OIdue$ (blue) line luminosities and stellar mass (left), gas-phase metallicity traced by oxygen-to-hydrogen number ratio with respect to solar (middle), and H$_2$ mass (right) for our simulated galaxies at $z = 6$. Solid lines represent the median trends, while density contours illustrate the underlying distribution of galaxies.}
    \label{fig:OIscaling}
\end{figure*}


\begin{figure*}
    \centering
    \includegraphics[width=0.9\columnwidth]{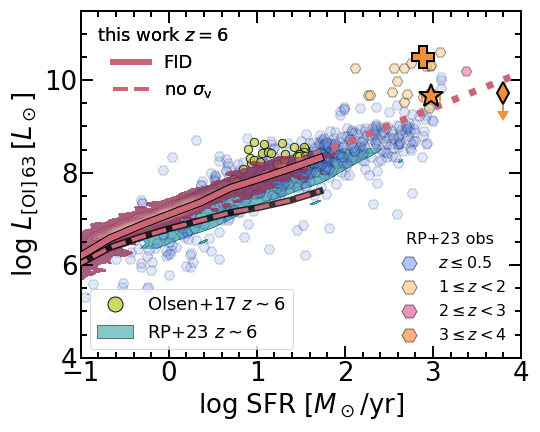}\quad
    \includegraphics[width=0.9\columnwidth]{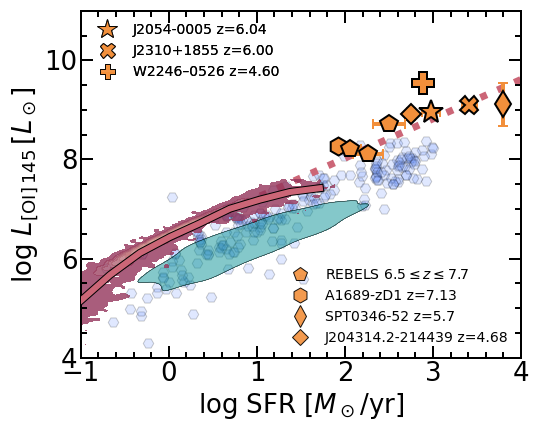}
    \caption{\small Relation between $\OI$ (left panel) and $\OIdue$ (right panel) line luminosity and SFR. Our simulation results at $z=6$ are shown as red density contours, with solid lines representing median trends. The median trend of the model with self-absorption neglecting the Eq. \ref{eq:velcond} condition is shown as a dashed line. Dotted lines are linear fits of the simulation results\rev{, with coefficients listed in Section \ref{sec:scaling}.}. Other $z \sim 6$ predictions for $\OI$ luminosity obtained from the post-processing of numerical simulations are also reported: \cite{Olsen17} as blue circles and \cite{RP23} as a turquoise area. Some observations are also reported for reference. The data up to $z=4$ compiled by \cite{RP23} are shown as hexagons. The J2054-00005 at $z=6.04$ \citep{Ishii2025} and W2246-0526 at $z=4.60$ \citep{FA24} have both $\OI$ and $\OIdue$ detections. As for $\OIdue$, we include the recent $6.5 \leq z \leq 7.7$ REBELS observations \citep[][]{Fudamoto25}, the $z=6.00$ J2310+1855 QSO \citep{Li2020}, and the new data obtained in this work (A1689-zD1 at $z=7.13$, SPT0346$-$52 at $z=5.7$ and J204313.2-214439 at $z=4.68$; see Sect. \ref{sec:obs}).}
    \label{fig:SFRoxy}
\end{figure*}

Here we present a set of scaling relations between the $\OI$ and $\OIdue$ line luminosities and various physical properties of our simulated galaxies at $z = 6$. The weak redshift evolution of these relations is discussed in App.~\ref{app:zevo}, while the predicted SFR–[CII] relation is in App.~\ref{app:CII}.\\
Figure \ref{fig:OIscaling} shows the $\OI$ and $\OIdue$ line luminosities of our simulated galaxies as functions of stellar mass, gas-phase metallicity $12+{\rm log(O/H)}$, and H$_2$ mass. We find a strong correlation between [OI] luminosities and both H$_2$ mass and stellar mass. As for stellar mass, the relation flattens at the high end, where the star-forming gas becomes predominantly cold ($T \lesssim 50 \, {\rm K}$). Such a drop is particularly evident for the $\OIdue$ line. Unsurprisingly, the correlation with H$_2$ mass is even tighter since both luminosities and H$_2$ are derived by DESPOTIC. However, the $\OI-M_{\rm H_2}$ relation features a change of slope at $M_{\rm H_2} \approx 10^8 \, M_\odot$, due to the impact of foreground self-absorption, which becomes relevant at approximately these molecular masses, which are associated with ${\rm SFR}\approx 1\, \rm M_\odot/{\rm yr}$ galaxies.
\\
In contrast, the relationship between $\OI$ and $\OIdue$ luminosities and gas-phase metallicity is considerably weaker, characterized by significant scatter. This increased dispersion is due to the implementation of metal production and diffusion in the simulation, which distributes metals among neighbouring gas particles. While this approach is included to roughly account for unresolved sub-grid mixing processes, it also introduces additional scatter when analyzing metallicity at the resolution of individual gas particles. The sharp transition around $12 + \log({\rm O/H}) \simeq 8.2$ marks the shift from galaxies dominated by warm gas ($T \gtrsim 100$ K) to those dominated by cold gas ($T \lesssim 100$ K), leading to a reduction in [OI] emission per unit gas mass, hence a sharp flattening in the luminosity-metallicity relation. \\
The relation between $\OI$ and $\OIdue$ luminosities and galaxy SFR is shown in Fig. \ref{fig:SFRoxy}. Alongside our simulated results at $z=6$, we include the results obtained from the post-processing of galaxy evolution simulations by \cite{Olsen17} and \cite{RP23}. Various observational results are also presented (Sect. \ref{sec:obs}). \\
The relation between our predicted luminosities and SFR is quite tight, with a mild flattening at the high-SFR end, attributed to the colder gas in these galaxies. This flattening is particularly evident for $\OIdue$, as its luminosity decreases more rapidly than that of $\OI$ with falling temperature. This is because the $^3 P_0$ level (from which the transition $^3 P_0 \rightarrow ^3 P _1$ originates) is hardly populated (as it requires $\Delta E/k \sim 325 \, {\rm K}$). 
We find good agreement between our predictions and those of other theoretical studies, which is particularly remarkable considering the different post-processing techniques and simulations employed. Notably, our simulation is the only one that accounts for foreground absorption of $\OI$ emission by cold gas, which reduces the total line luminosity by a factor of approximately $\sim 2$–$4$ (Fig. \ref{fig:betaOI}). The self-absorption model without the relative velocity condition (Eq.~\ref{eq:velcond}) shows less agreement at high SFRs, where the effect is likely overestimated ($\approx 90 \, \%$ of the flux is self-absorbed).
\\
Due to the limited volume of our simulation, we cannot sample highly star-forming systems ($\gtrsim 100 \, M_\odot/{\rm yr}$), which are predominantly the ones observed at high redshift. 
When comparing our extrapolated relations with observational data, we find good agreement, except for the W2246–0526 $\OI$, which lies about an order of magnitude above our extrapolation. 
Note, however, the complex nature of this object, which is the most luminous galaxy in the Universe, likely hosting a heavily dust-obscured QSO \citep{DiazSantos16}.
The extrapolation is obtained by fitting our results with the relation
$\log (L_{\rm [OI]} / L_\odot ) = a \, \log ({\rm SFR} / {\rm M_\odot yr^{-1}}) + b$, 
with coefficients $(a_{\OI}, b_{\OI}) = (0.76, 7.11)$ and $(a_{\OIdue}, b_{\OIdue}) = (0.89, 6.40)$ for $ \OI$ and $\OIdue$, respectively. \rev{The scatter of the relation is $\approx 0.5 \, {\rm dex}$ for both lines and almost SFR--independent.}

\subsection{Line ratios}
\label{sec:ratio}

\begin{figure}[htb!]
    \centering

    \begin{subfigure}{\linewidth}
        \centering
        \includegraphics[width=0.9\linewidth]{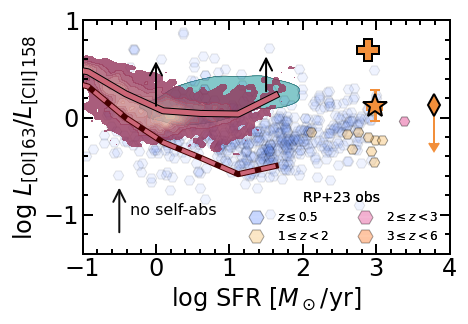}
        \label{fig:ratio:63158}
    \end{subfigure}

    \vspace{0.2em} 

    \begin{subfigure}{\linewidth}
        \centering
        \includegraphics[width=0.9\linewidth]{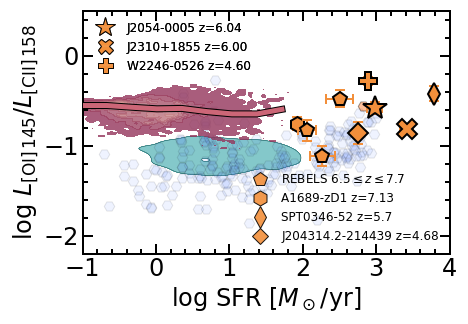}
        \label{fig:ratio:145158}
    \end{subfigure}

    \vspace{1em}

    \begin{subfigure}{\linewidth}
        \centering
        \includegraphics[width=0.9\linewidth]{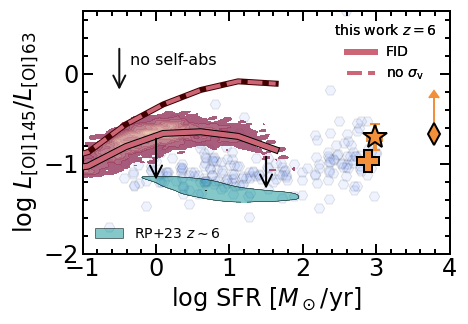}
        \label{fig:ratio:14563}
    \end{subfigure}

    \caption{\small Lines ratio as a function of SFR: $\OI / \CII$ (top), $\OIdue / \CII$ (middle) and $\OIdue / \OI$ (bottom). Our simulation results at $z=6$ are shown with red contours, with the solid line representing median trend. The median trend of the model with self-absorption neglecting the Eq. \ref{eq:velcond} condition is shown as a dashed line. The black arrows indicate the magnitude of the $\OI$ self-absorption by foreground gas as predicted by our model. Simulation predictions by \cite{RP23} at $z \sim 6$ are shown as turquoise contours. The same observational data shown in Fig. \ref{fig:SFRoxy} are included where detections are available, as described in the legend.}
    \label{fig:ratios}
\end{figure}

In Fig. \ref{fig:ratios} we report our model predictions for line ratios at $z=6$ (i.e. $\OI/\CII$, $\OIdue/\CII$ and $\OIdue/\OI$) and compare them with available observations.
We note that the $\OI/\CII$ and $\OIdue/\CII$ ratios increase by a factor of $\approx 2-3$ at higher redshifts, 
whereas the $\OIdue/\OI$ ratio shows little redshift evolution (see App.~\ref{app:zevo} for details).
We note that our simulation probes 
normal star forming galaxies with 
SFRs up to $\approx 100 \, {\rm M_\odot /  yr}$, generally below the SFRs of powerful high-$z$ observed sources. Therefore, 
extrapolations above those values must be taken with caution.

\subsubsection{$\OI / \CII$}


The top panel of Fig. \ref{fig:ratios} shows the $\OI/\CII$ ratio. Our model predicts a ratio of $\approx 1$, slightly increasing at both low and high SFR (up to $\approx2$). The low SFR end increase is due to the lower contribution of foreground $\OI$ self-absorption, while the high end one is mainly driven by the flattening of the $\CII -$SFR relation (App. \ref{app:CII}). Observations suggest a $z-$independent ratio close to $1$, with a factor of $\approx 2$ scatter. The ratio measured for J2054-0005 is consistent with our values, whereas W2246-0526 is somewhat higher; however, both objects have SFRs significantly above those of our simulated galaxies.
Our results incorporate the effect of $\OI$ self-absorption by cold foreground gas. Without this contribution, the $\OI/\CII$ ratio would be a factor of $\approx 2$ higher, slightly overestimating the observational data. We note that the simulations-based ratios reported by \cite{RP23}, where this effect is not taken into account, are indeed slightly larger than ours.\\
Finally, we note that the self-absorption model without the condition of Eq. \ref{eq:velcond} yields ratios well below the observed high-$z$ values (with the exception of the SPT0346-52 upper limit).
Interestingly, De Breuck et al. (in prep.) report $\OI/\CII$ ratios reaching $\lesssim 0.01$ in a sample of 12 SPT high$-z$ dusty star-forming galaxies, suggesting a stronger role of self-absorption and leaving room for possible refinements of this modeling in the high-SFR regime.

\subsubsection{$\OIdue / \CII$}


The $\OIdue/\CII$ ratio offers a potential advantage over the $\OI/\CII$ ratio, as the $145\,\mu\mathrm{m}$ oxygen line is optically thin and thus not affected by uncertainties from foreground self-absorption. Moreover, this ratio is particularly useful in observations for inferring gas densities \citep{Peng25b}.
We include recent results from \cite{Fudamoto25} for $z \sim 7$ REBELS galaxies and the $z = 6.00$ QSO J2310+1855 \citep{Ishii2025}, along with new observations from this work presented in Sect. \ref{sec:obs:newdata}, which nearly double the number of available data points for this ratio at such high redshift.
\\
Our model predicts a constant $\OIdue / \CII $ value of $ \approx 0.15$ with a scatter of a factor $\lesssim 2$ (Fig. \ref{fig:ratios}, central panel). Our estimates are a factor of $\approx 2$ higher than those presented by \cite{RP23}.
Despite the large scatter, observations at $z \gtrsim 5$ show $\OIdue/\CII$ ratios ranging from $0.08$ to $0.4$.
Although simulated SFRs do not exactly overlap with the observed SFR range, observed ratios are fully consistent with our simulation values.
No significant variation is found among the QSOs in the sample.

\subsubsection{$\OIdue / \OI$}

The $\OIdue / \OI$ ratio is a great indicator of the self-absorption of $\OI$, since the two lines have similar critical densities and $\OIdue$ being optically thin. Without self-absorption effects, this ratio is expected to be $\lesssim 0.1$ for gas with typical physical ISM conditions (see discussion in Sect. \ref{sec:disc:sa}). 
Due to self-absorption, our model galaxies feature a $\OIdue / \OI$ ratio of $\approx  0.25-0.1$, instead of being a factor of $2-4$ lower (Fig. \ref{fig:ratios}, bottom panel). The ratio features a decrease at both low and high SFRs, respectively due to the lower impact of $\OI$ and the high-SFR flattening of the $\OIdue -$SFR relation. Note that the self-absorption factor is approximately the gap with \cite{RP23} predictions, which do not account for this effect. Our model predicts slightly higher values than observed at $z \lesssim 0.5$, but yields ratios consistent with the only detections currently available at $z \gtrsim 4.6$ (J2024-0005 and W2246-0526). Instead, when using the self-absorption model that neglects the condition in Eq. \ref{eq:velcond}, values up to $\approx 1$ are obtained in the high-SFR regime. 

\section{Discussion}
\label{sec:discussion}

\subsection{Foreground $\OI$ self-absorption}

\label{sec:disc:sa}

\begin{figure}
    \centering
    \includegraphics[width=1\columnwidth]{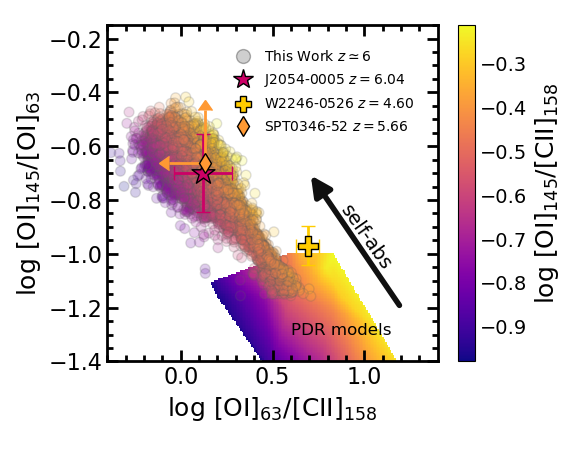}
    \caption{\small Distribution of the $\OIdue / \OI$ and $\OI / \CII$ line ratios for different values of $\OIdue / \CII$, as shown in the colorbar. The colored region represents the outcome of the PDR DESPOTIC computations assuming homogeneous clouds with $10^2 \leq n_{\rm H}/{\rm cm}^{-3}\leq 10^6$ and $20 \leq T/{\rm K} \leq 10^4$. Our simulated $z \simeq 6$ galaxies are shown as circles, J2054-0005 as a star, W2246-0526 as a plus and the SPT0346$-$52 upper limit as a diamond. The black arrow illustrates the typical shift due to $\OI$ foreground self-absorption.}
    \label{fig:contour_ratio}
\end{figure}

Our results highlight the relevance of cold foreground self-absorption of the $\OI$ line. This reduces the observed intensity by a factor of $2-4$, which is consistent with the approach adopted by various observational works. However, being this value very uncertain studying the ratio between $\OI$ and other optically thin lines ($\OIdue$, $\CII$) becomes essential to investigate the impact of self-absorption. Our results, together with the observations available (Fig. \ref{fig:ratios}, central panel), indicate a $\OIdue / \CII$ ratio of $\approx 0.1$-0.6. 
We conduct a set of experiments with DESPOTIC to see which are the values of $\OIdue / \OI$ and $\OI / \CII$ expected to be consistent with the above ratio. In these experiments, we assume homogeneous clouds and the standard GOW chemical network (Sect. \ref{sec:mod:lines}), letting the gas density and temperature varying in the range $n_{\rm H} \in [10^2, 10^6] \, {\rm cm}^{-3}$ and $T \in [20, 10^4]\, {\rm K}$. A radius of $10 \, {\rm pc}$ is assumed to convert the volume density to column density. Results of this experiments are shown in Fig. \ref{fig:contour_ratio} as a region colored according to the $\OIdue / \CII$ ratio. In the same figure we also report our simulated galaxies together with J2054-0005, W2246-0526 and SPT0346$-$52 (limits). 
The typical densities and temperatures needed to produce the observed ratio of $\OIdue / \CII$ would also produce a $\OI / \CII$ and $\OIdue / \OI$ ratios of, respectively, $\approx 2-10$ and $\lesssim 0.1$. These values differ by a factor of about $2$ from those measured in J2054-0005. A similar, or even larger, discrepancy is suggested by the upper limits of SPT0346$-$52. In contrast, W2246–0546 is closer to these values.
\\
Our simulated galaxies -- for which foreground $\OI$ self-absorption is taken into account -- are instead more in line with observations. This simple yet motivated reasoning highlights the need of $\OI$ observations (or even non-detections) of targets with known $\OIdue$ and $\CII$ fluxes, which would constrain the foreground self-absorption at such high $z$.
\\
We conclude by noting that \cite{Peng25a}, who recently compiled an extensive dataset of FIR emission lines across a wide redshift range, also highlighted challenges in reproducing the observed $\OIdue/\OI$ and $\OIdue/\CII$ line ratios with theoretical models. They suggested that $\OI$ self-absorption could be a significant factor contributing to these discrepancies.

\subsection{Mass-to-Light ratios $M_{\rm H_2}/L_{\rm [OI]}$}


\begin{figure}
    \centering
    \includegraphics[width=0.8\columnwidth]{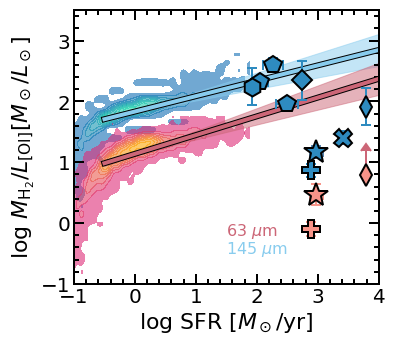}
    \caption{\small Mass-to-Light ratio $M_{\rm H_2}/L_{\rm [OI]}$ for both $\OI$ and $\OIdue$, respectively in red and blue color scales. We report simulation predictions at $z = 6$ (contours) and linear fits for log(SFR$/{M_\odot {\rm yr}^{-1}}$)$\geq - 0.5 $. Observational results (Sect. \ref{sec:obs}) for REBELS galaxies (pentagons), A1689-zD1 (hexagon), J204313.2-214439 (diamond), SPT0346$-$52 (small diamond), W2246-0526 (plus), J2054-0005 (star), and J2310+1855 (cross) are also reported.}
    \label{fig:mtl}
\end{figure}

We discuss our $z=6$ model predictions for the mass-to-light ratio $M_{\rm H_2}/L_{\rm [OI]}$ as a function of SFR and compare them with the observations available in Fig.~\ref{fig:mtl}. 
Our simulated galaxies feature a slowly increasing $M_{\rm H_2}/L_{\rm [OI]}-{\rm SFR}$ relation at SFR$\gtrsim 0.3 \, {\rm M_\odot / {\rm yr}}$, and the slope is similar for the two lines. Below this threshold, the ratio rapidly increases, due to the scarcity of molecular mass in galaxies with typically low column densities ($N_{\rm H} \lesssim 10^{22} \, {\rm cm}^{-2}$). Excluding these points, we fit our results with a simple linear relation of the form $\log \left( \frac{L_{\rm [OI]}}{M_{\rm H_2}} \cdot \frac{M_\odot}{L_\odot} \right) = a \cdot \log \left( \frac{\rm SFR}{M_\odot,{\rm yr}^{-1}} \right) + b$, with the resulting coefficients listed in Table \ref{tab:coefffit}, along with those obtained from fits as a function of metallicity and stellar mass.
%
%
\begin{table}[]
\centering
\caption{\small Best-fit parameters for the fit relations $\log y = a \, \log x + b$, with $y = L_{\rm [OI]}/M_{\rm H_2}$ in solar units and $x$ SFR, metallicity, or stellar mass.}
\label{tab:coefficients}
\begin{tabular}{lcccc}
\hline
$x$ & Line & $a$ & $b$ \\
\hline
\multirow{2}{*}{$\mathrm{SFR\,[M_\odot \, yr^{-1}]}$} 
& $\OIdue$ & $0.253 \pm 0.002$ & $1.834 \pm 0.001$ \\
& $\OI$  & $0.308 \pm 0.004$ & $1.139 \pm 0.002$ \\
\hline
\multirow{2}{*}{$12 + \log(\mathrm{O/H})$} 
& $\OIdue$ & $0.69 \pm 0.01$ & $-3.6 \pm 0.1$ \\
& $\OI$  & $0.90 \pm 0.01$ & $-6.0 \pm 0.1$ \\
\hline
\multirow{2}{*}{$M_{\rm stars}\,[\rm M_\odot] $} 
& $\OIdue$ & $0.262 \pm 0.002$ & $-0.35 \pm 0.02$ \\
& $\OI$  & $0.257 \pm 0.004$ & $-0.98 \pm 0.04$ \\
\hline
\end{tabular}
\label{tab:coefffit}
\end{table}
\\
Two main factors drive the slow increase in the mass-to-light ratio observed in all three cases. First, the gas in star-forming, massive, and metal-rich galaxies tends to be colder, which lowers the specific luminosity of emission lines. Second, the higher metallicity -- associated with increased dust content in our post-processing model -- combined with the high column densities typical of more massive galaxies, creates optimal conditions for H$_2$ formation.
\\
Our model well reproduces the $\approx 0.003-0.01$ $L_{\OIdue}/M_{\rm H_2}$ ratio featured by the REBELS galaxies and A1689-zD1, although these have a larger scatter than our tight relation. Instead, when comparing to the bright SMGs (SFR $\gtrsim 1000\, M_\odot / {\rm yr}$) and QSOs, our predictions are above observationally derived ratios up to one order of magnitude, for both the $\OI$ and $\OIdue$ line. This implies typical lower mass-to-light ratios for the SMG and QSO population, as already noticed for the $\CII$ line \citep{Salvestrini25}, and in general for highly star forming galaxies \citep{Rizzo20, Rizzo21}.
\\
Finally, we note that although the H$_2$ abundance predicted by the simulation is used within the star formation model, in this work we refer to molecular gas masses derived using DESPOTIC. 
The inferred estimates are valid for sub-grid unresolved scales and are typically larger by $\sim$1 dex compared to the on-the-fly values obtained for resolved scales.
We choose to adopt the post-process values, because they are discussed in the context of line luminosities, which are also computed in post-processing and account for sub-resolution structures at densities higher than those resolved in the simulation.

\subsection{Model variations}
\label{sec:modvar}

\begin{figure}
    \centering
    \includegraphics[width=0.9\columnwidth]{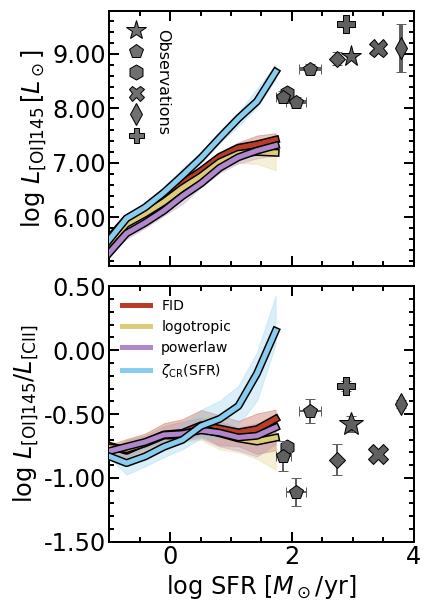}
    \caption{\small $\OIdue$ luminosity and $\OIdue / \CII$ ratio as a function of SFR for our model galaxies at $z=6$. Different colors refer to different model assumption in the post-processing pipeline with DESPOTIC. The red one represents the fiducial model adopted in this work, the cyan is a model where the CRs rate is scaled with the SFR density, the yellow and purple are models where the density profile is assumed to be logotropic and power-law. Solid lines refer to median trends and shaded regions indicate the $16-84$th-percentile dispersion. For reference, the available $z \gtrsim 5 $ observations discussed in this work are also reported, with the same symbols as in Fig. \ref{fig:SFRoxy} and \ref{fig:ratios}.}
    \label{fig:moreflavs}
\end{figure}

In this section, we present a series of numerical tests using different DESPOTIC configurations to assess the sensitivity of our results to key model assumptions. Specifically, we examine the impact of varying the CR heating prescription and the internal density profile of molecular clouds.
Figure \ref{fig:moreflavs} shows the resulting $\OIdue$ luminosities and $\OIdue / \CII$ ratios as functions of the SFR for our $z = 6$ model galaxies under different parameter choices. Although we focus here on a subset of lines, we checked that similar trends are observed also in other emission lines and ratios.
\\
We begin by examining the impact of varying the cloud density profiles. Adopting either a logotropic or power-law profile \citep{Popping19}
-- both commonly used in the literature \citep[e.g.,][]{Olsen17, Khatri2024, Garcia2024} -- results in only minor changes to the predicted emission. This finding confirms the results of \cite{Popping19}, who explored variations in density profiles and other sub-resolution cloud properties. The limited impact is essentially due to the fact that both the logotropic and power-law profiles converge toward the average density in the outer regions of the clouds, which dominate the emission due to their higher temperatures. As a result, the total line luminosities remain largely unaffected.
\\
Next, we modify the CR heating model in DESPOTIC by implementing a scaling with the (unattenuated) SFR density, i.e. $\zeta_{\rm CR} = 10^{-17} \cdot \Sigma^{\rm unatt}_{\rm SFR}/\Sigma^{\rm unatt}_{\rm SFR, \, MW}$. This prescription is widely adopted in the literature (e.g., \citealt{Popping19}; \citealt{Garcia2024}). As anticipated by the work of \cite{Krumholz2023}, which motivates our CRs treatment, this SFR-based scaling leads to CRs energy densities also $2-3$ orders of magnitude higher in the most massive galaxies compared to our fiducial model. Consequently, the gas in GMCs becomes significantly warmer, with typical temperatures around $T_{\rm gas} \approx 100\,{\rm K}$, compared to $\approx 30\,{\rm K}$ in the fiducial case. This increased heating boosts line emission in strongly star-forming systems and prevents the suppression of $\CII$ and $\OIdue$ luminosities. However, while this improves agreement with observed line luminosities, it leads to $\OIdue / \CII$ and $\OI / \CII$ ratios that are too high, typically exceeding the observed values of $\approx 1$ and $\approx 0.3$~dex, respectively.
\\
One possible explanation for this discrepancy might be a missing contribution to $\CII$ emission from warm/hot ionized gas. At temperatures $T \gtrsim 10^4\,{\rm K}$, oxygen is expected to be ionized, while carbon can remain singly ionized up to a few $10^4\,\rm K$.
To account for this, we include a $\CII$ contribution from warm, non-star-forming gas particles by using the density and temperature predicted by our hydrodynamic simulation. We restrict this analysis to particles with $T \leq 10^5\,{\rm K}$, under the (optimistic) assumption that CIII dominates at higher temperatures. The resulting contribution is negligible, though, and typically accounts for only $\approx 0.01 - 0.1$ of the total $\CII$ luminosity, due to the low densities involved ($n_{\rm H} \lesssim 1\,{\rm cm}^{-3}$).
To test whether this result is resolution-limited, we also model the density of the warm/hot gas as a log-normal distribution. However, this yields similar results, confirming the limited role of hot-phase $\CII$ emission.

\subsection{Comparison with literature}

We briefly discuss our results in the context of previous simulations and observations, as well as the new observations analyzed in this study.\\
%
%
Previous hydrodynamical simulations at $z \simeq 6$ \citep[e.g.,][]{Olsen17, RP23}, have attempted to model [OI] lines and ratios with more simplified treatments.
A key difference lies in our inclusion of $\OI$ self-absorption due to foreground gas, a physical effect that has not been accounted for in other works. Although $\OI$ luminosities in \citet{RP23} are comparable to ours, their $\OIdue/\CII$ ratios are systematically lower. This suggests that their model underestimates the intrinsic emission strength of both [OI] lines. We point out that, instead, our inclusion of self-absorption allows our model to better reproduce the full set of observed ratios.
\\
%
%
A comparison of our predictions with both the new data presented here and the observations available in the literature \citep[][]{Ishii2025, Fudamoto25, Li2020} reveals broadly consistent trends in $\OI, $$\OIdue$, and $\CII$ line luminosities as a function of SFR.
Predicted and observed ratios among different line tracers are also consistent.
These data, together with the compilation by \citet{RP23}, do not show a strong dependence of these quantities on redshift. This supports the idea of weak or negligible evolution in the luminosity–SFR relation and line ratios, in agreement with our findings over $6 \leq z \leq 10$. \rev{We remark that the high-$z$ observations typically span a higher SFR regime than that covered by our simulations. Deeper observations of $z \approx 6$ targets with SFRs $\lesssim 100 \, \rm M_\odot \, yr^{-1}$ are challenging with current facilities, as this would require exposure times of several hours per galaxy. 
Alternatively, applying a post-processing pipeline like the one presented here may enable comparisons with high-$z$ systems in large-volume cosmological simulations -- e.g. MillenniumTNG \citep{MillenniumTNG}; Magneticum Pathfinder \citep{Dolag25}; COLIBRE \citep{COLIBRE} -- which can host galaxies with SFRs up to $\approx 10^3 \, \rm M_\odot \, yr^{-1}$ at $z \approx 6$.}
\\
\rev{While in this work we focus on simulated star formation–powered systems, the radiative contribution from AGNs may play a role in the resulting line emission. 
Previous studies reported higher $\OI / \CII$ ratios in AGN-dominated systems due to the enhanced cooling efficiency of $\OI$ in warm, dense gas \citep{HC18} or the highly ionizing radiation from AGNs \citep{Li2020}. This is also consistent with the $\CII /$FIR deficit observed in AGNs \citep[e.g.][]{Venemans20}.}\\
Finally, in the three lensed galaxies and one QSO host galaxy at $z \gtrsim 4.7$ analyzed in this work, we measure $\OIdue$ luminosities $\gtrsim 2\times10^8$ $L_\odot$. These values are at the bright end of the ones previously reported by \cite{Novak19} and \cite{Fudamoto25} and make these sources among the largest molecular-gas reservoirs at these epochs (Table \ref{tab:physpropobs}).

\section{Conclusions}
\label{sec:conclusions}
In this paper, we performed theoretical predictions for the line emission of atomic neutral oxygen ($\OI$ and $\OIdue$) in high-redshift galaxies. We exploited the \coldsim cosmological hydrodynamic simulations post-processed with the radiative transfer code \textsc{DESPOTIC} to model the chemical, thermal and statistical state of sub-resolution ISM clouds and compute the resulting line luminosities. 
For the first time, our pipeline includes a physically motivated model for $\OI$ self-absorption by cold foreground gas, an effect often neglected in previous theoretical works.
We also presented new ALMA measurements of $\OIdue$ emission in four objects at $z \gtrsim 5$.
\\
Our main findings can be summarized as follows.
\begin{itemize}
    \item We find strong correlations between SFR and $\OI$ or $\OIdue$ luminosities, while diagnostic line ratios, such as $\OI/\CII$, $\OIdue/\CII$, and $\OIdue/\OI$, are nearly independent of SFR. These relations show minimal evolution over $6 \leq z \leq 10$.
    
    \item ALMA observations of $\OIdue$ show luminosities in the range $\sim 10^8$-$10^9\,\rm L_\odot$, a factor $\sim$4-7 lower than [CII] luminosities, and inferred ratios $\OIdue$/$\OI \gtrsim 0.2$.

    \item Our model of $\OI$ self-absorption shows that intrinsic $\OI$ luminosities are about 2-4 times higher than what is actually observed and modeling self-absorption is essential to reproduce the observed line ratios successfully.

    
    \item Both $\OI$ and $\OIdue$ lines are good tracers of molecular gas in normal star forming galaxies and this makes them valuable alternatives to traditional tracers like CO, which is affected by the CMB at high $z$ \citep{daCunha13}. Current measurements of $M_{\rm H_2}$-to-$L_{[\rm OI]}$ ratios in highly star forming galaxies and QSOs fall below our predictions, possibly indicating a deficit of molecular content, different ISM conditions or extreme excitation mechanisms in those environments \citep{Bischetti21, Circosta}.
\end{itemize}

\noindent
This work is relevant for the relatively new subject of self-absorbed high-$z$ fine-structure lines and establishes the way for further scientific developments in the field.
Our results, in keeping with observations, offer robust predictions for future detections and support the use of [OI] lines as tracers of star formation activity and molecular gas in the early Universe. Despite the effects of self-absorption, the $\OI$ line remains as bright as, or even brighter than, the $\CII$ line. However, its detectability at $z \gtrsim 6$ is limited by atmospheric opacity in ALMA Bands 8–10. In contrast, the lower optical depth of the $\OIdue$ line makes it more accessible to ALMA,
as indicated by the high luminosities measured for the four high-$z$ galaxies considered in this work.
Upcoming ALMA observations targeting the two [OI]~lines discussed here will provide valuable constraints on the physical conditions of the ISM and the role of $\OI$ self-absorption during the epoch of reionization.

\begin{acknowledgements}
\rev{We thank the anonymous referee for the useful suggestions which improved this manuscript, and the editor DE for the support during the publication process. MP acknowledges useful discussions with D. Narayanan, K. Garcia, and D. Valentin-Martinez, and thanks the organizers of the Infrared Fine-Structure Lines Workshop 2025 at Winona State University.}
This paper makes use of the following ALMA data: ADS/JAO.ALMA\#2023.1.01450.S (P.I. C.~Ferkinhoff) and ADS/JAO.ALMA\#2021.1.00265.S (P.I. D. Riechers). ALMA is a partnership of ESO (representing its member states), NSF (USA), and NINS (Japan), together with NRC (Canada), MOST and ASIAA (Taiwan), and KASI (Republic of Korea), in cooperation with the Republic of Chile. The Joint ALMA Observatory is operated by ESO, AUI/NRAO and NAOJ.
UM and MP acknowledge financial support from grant no. 1.05.23.06.13 “FIRST -- First Galaxies in the Cosmic Dawn and the Epoch of Reionization with High Resolution Numerical Simulations”, awarded by the Italian National Institute of Astrophysics (INAF). UM also acknowledges financial support from INAF travel grant no. 1.05.23.04.01. CR and GLG acknowledge financial support from the European Union's HORIZON-MSCA-2021-SE-01 Research and Innovation Programme under the Marie Sklodowska-Curie grant agreement number 101086388 - Project (LACEGAL).
CF, FS and MB acknowledge financial support from Ricerca Fondamentale INAF 2023 Data Analysis grant 1.05.23.03.04 ``ARCHIE ARchive Cosmic HI \& ISM  Evolution''. FS and MB acknowledge financial support from the PRIN MUR 2022 2022TKPB2P - BIG-z, Ricerca Fondamentale INAF 2024 under project 1.05.24.07.01 MINI-GRANTS RSN1 "ECHOS", Bando Finanziamento ASI CI-UCO-DSR-2022-43 CUP:C93C25004260005 project ``IBISCO: feedback and obscuration in local AGN''.
\end{acknowledgements}


\bibliographystyle{aa} 
\bibliography{lines} 


\clearpage
\newpage

\appendix

\section{RF accuracy}

\label{app:RF}
\begin{figure*}
    \centering
    \includegraphics[width=0.6\columnwidth]{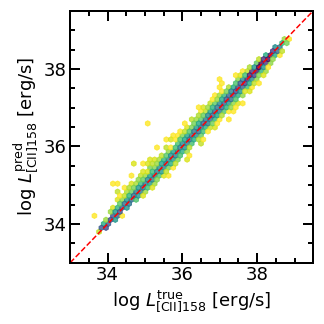}\quad
    \includegraphics[width=0.6\columnwidth]{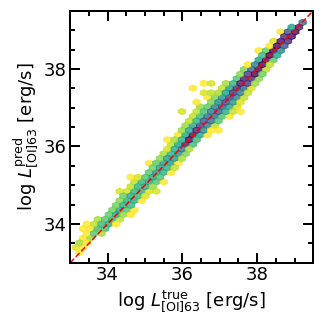}\quad
    \includegraphics[width=0.6\columnwidth]{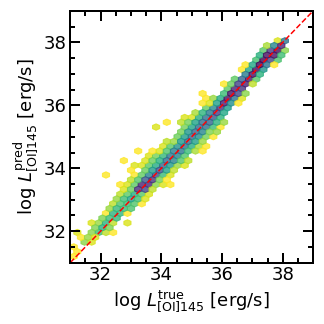}
    \caption{\small Relation between the true luminosity (x-axis) and RF predicted luminosity (y-axis) for the particles ($20 \%$ of the total sample of $\sim 5 \cdot 10^5$) adopted for testing the performances of the trained RF algorithm. Results for the three predicted luminosities, i.e. [CII]$_{158 \mum}$, [OI]$_{63 \mum}$ and [OI]$_{145 \mum}$, are shown respectively in the left, middle and right panel.}
    \label{fig:MLtest}
\end{figure*}

We briefly check the performances and accuracy of the RF algorithm described in Sect. \ref{sec:ML} adopted to compute the lines luminosity of the gas particles starting from the $5$ parameters provided by the cosmological simulation. The relation between the true and predicted luminosity of the test particles adopted by the RF algorithm is shown in Figure \ref{fig:MLtest} for the $\CII$, $\OI$ and $\OIdue$ lines. The accuracy of the RF is quantified by a coefficient of determination $R^2$ defined as
\begin{equation}
    R^2 = 1 - \frac{\sum_i (y_{i, \rm true}-y_{i, \rm pred})^2}{\sum_i (y_{i, \rm true}-\bar{y}_{\rm true})^2},
\end{equation}
with the subscripts true and pred referring to the true and predicted values. For the model presented here $R^2 \approx 0.996$.


\section{ALMA maps of [OI] 145$\mu$m, [CII], and continuum}\label{app:newdata}

\begin{table}[]
    \setlength{\tabcolsep}{3pt}
    \caption{\small New $\OIdue$ observations considered in this work. 
    }
    \label{tab:almadata}
    \centering
    \begin{tabular}{lccc}
    
    \hline
    ID & beam & rms$_{30}$  & Flux \\
       & arcsec$^2$ & mJy beam$^{-1}$ & Jy \kms\\
    \hline
    A1689$-$zD1 & 1.41$\times$1.18 & 0.22 & 0.567$\pm$ 0.058 \\

    SPT0346$-$52 & 1.10$\times0.65$  & 0.45 & $7.28 \pm 0.70$  \\


    J204314$-$214439 & 0.75$\times$0.65 & 0.42 & $3.28\pm0.30$\\
    

    J2054$-$0005 & 1.18$\times$0.96 & 0.23 & 0.847$\pm$0.028\\

    \hline
    \end{tabular}    
    \vspace{-0.5cm}
\end{table}

\begin{figure}[h!]
    \centering
    \includegraphics[width=0.49\columnwidth]{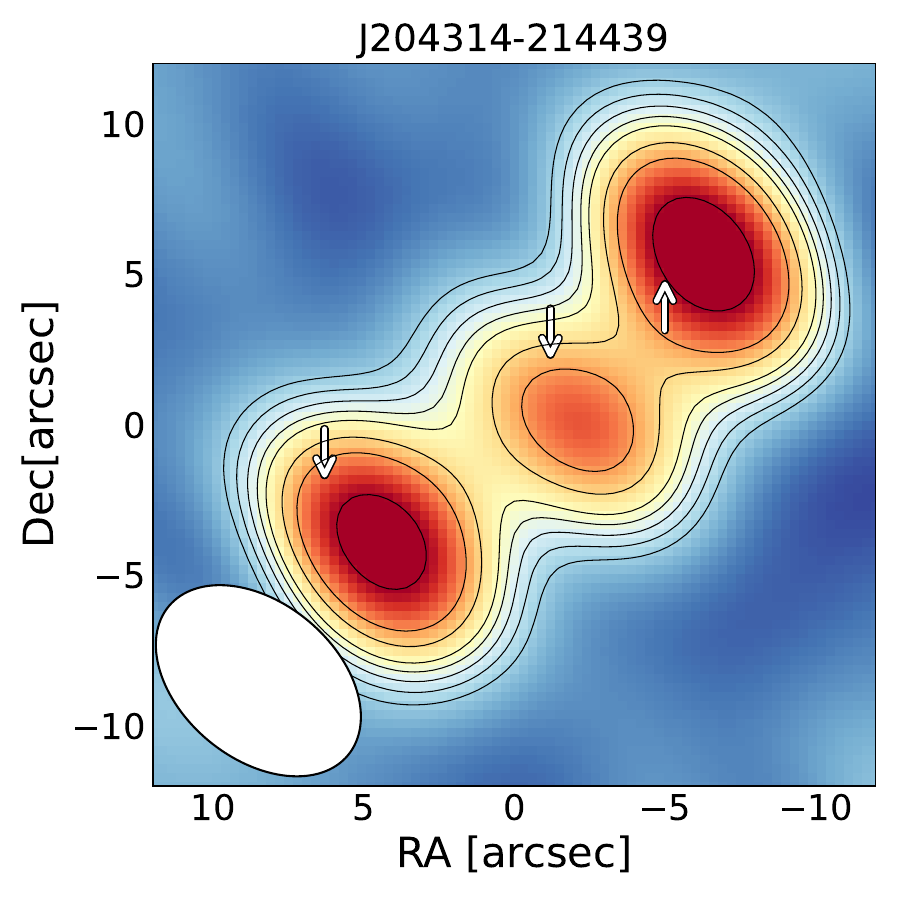}
    \includegraphics[width=0.49\columnwidth]{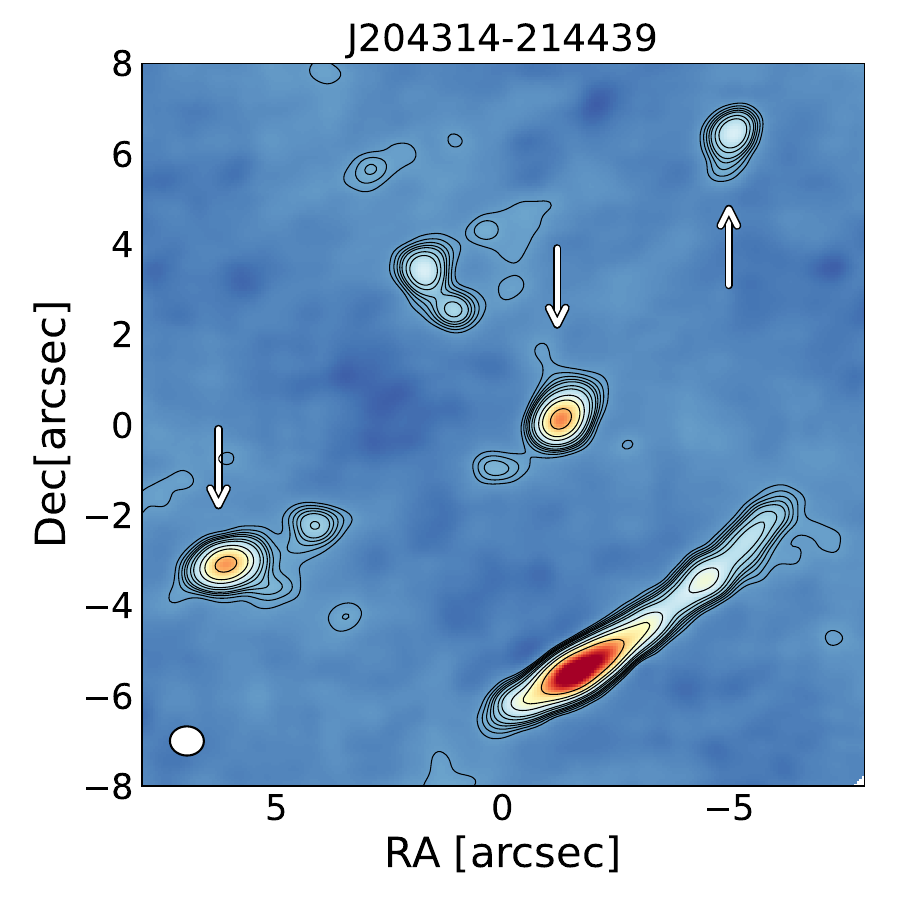}
    \caption{\small Left panel: [CII] emission map in J204314$-$214439. Contours correspond to [2, 3, 4, 5, 6, 8, 10, 15, 20, 30]$\,\sigma$ confidence levels, where $\sigma=0.58$~Jy~beam$^{-1}$~km ~s$^{-1}$. The bottom-left white ellipse shows the ACA beam. Right panel: Continuum map at $362.7$~GHz. Contours are as in left panel with $\sigma=0.12$~mJy ~beam$^{-1}$. Arrows indicate continuum sources in which we detect [OI] and [CII] emission. 
    The bottom-left white ellipse shows the ALMA beam.
    }
    \label{fig:hlsmaps}
\end{figure}

This section summarizes the analysis of the new ALMA data considered  in this work (Sect. \ref{sec:obs}).
Visibilities were calibrated using the standard calibration provided by the ALMA observatory and the default phase, bandpass and flux calibrators. To model and subtract the continuum emission from the line, we combined the adjacent spectral windows in the baseband containing [OI] or [CII] and performed a fit in the $uv$ plane to channels with $|v|>700$ \kms, using a first-order polynomial continuum. A continuum-subtracted datacube was created using CASA task $tclean$, with the $hogbom$ cleaning algorithm in non-interactive mode, a threshold equal to two times the rms sensitivity and a natural weighting of the visibilities. We adopted a 30 \kms\ channel width. The resulting synthesized beam and rms sensitivity for a 30 \kms\ channel are listed in Table \ref{tab:almadata}. We created 0$^{th}$ moment maps by integrating over the emission line channels. Velocity-integrated line intensities were measured from 2D Gaussian fit of the 0$^{th}$ moment maps (Table \ref{tab:almadata}).
\\
We created a continuum-subtracted ACA datacube of the [CII] emission in J204314$-$214439 following the approach described above, with an improved angular resolution (7.3$\times$5.2 arcsec$^2$) with respect to previous LMT observations \citep{Zavala2015} following the same approach as described above. The resulting [CII] O$^{th}$ moment map is shown in Fig. \ref{fig:hlsmaps} left. We detect three images with a similar [CII] line profile, two of which correspond to the double [OI] images detected in the ALMA map (Figure \ref{fig:newoi}c). The velocity-integrated [CII] flux has been obtained by summing the three images is $S_{\rm [CII]}=25.62\pm1.07$ Jy \kms.
\\
A high resolution map (0.75$\times$0.65 arcsec$^2$) of the $\sim362.7$ GHz continuum in J204314$-$214439 was created by averaging the ALMA visibilities over all spectral windows ($\sim3$ GHz) and excluding the channels of the [OI] emission, as shown in Figure \ref{fig:hlsmaps} right. By summing the fluxes of the three continuum sources in which we also detect $\CII$ and $\OIdue$ emission, we measure $S_{\rm 3.7}=24.47\pm1.17$ mJy.

\section{Redshift evolution of scaling relations}
\label{app:zevo}

\begin{figure*}
    \centering
    \includegraphics[width=0.6\columnwidth]{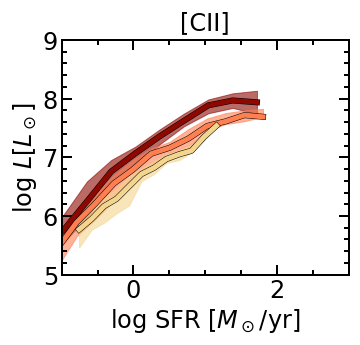}\quad
    \includegraphics[width=0.6\columnwidth]{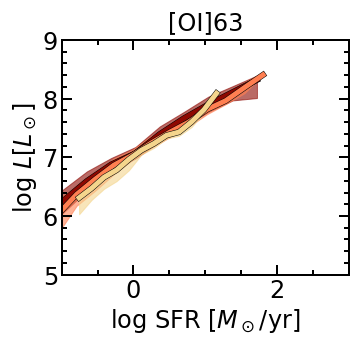}\quad
    \includegraphics[width=0.6\columnwidth]{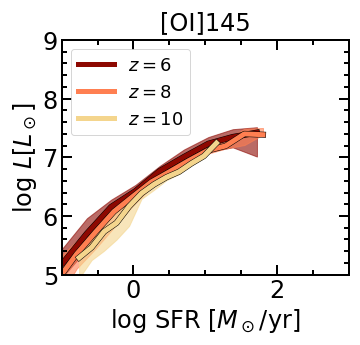}
    \caption{\small Relation between SFR and $\CII$, $\OI$ and $\OIdue$ luminosity for our simulated galaxies at $z=6, \, 8, \,{\rm and} \,10$. Solid lines refer to median trends, while shaded regions are $16-84$th-percentile dispersion.}
    \label{fig:zevo}
\end{figure*}

Fig. \ref{fig:zevo} shows the evolution of the SFR–luminosity relations for our simulated galaxies at redshifts $z = 6$, $8$, and $10$. We find that the $\OI$ and $\OIdue$ luminosities exhibit little to no redshift evolution, while the normalization of the SFR–[CII] relation increases slightly toward lower redshift. This trend is due to two competing effects: as redshift decreases, galaxies become more metal-enriched~-- particularly in carbon and oxygen~-- boosting line emission. Conversely, higher-redshift galaxies with similar SFRs tend to have higher gas densities. For [CII], metallicity enrichment dominates, leading to a modest redshift evolution in its luminosity. In contrast, the [OI] lines, which have higher critical densities, are more sensitive to gas density; thus, the two effects roughly balance, resulting in nearly redshift-invariant [OI] luminosities.
\\
Consequently, also the line ratios $\OI/\CII$ and $\OIdue/\CII$ exhibit a redshift dependence, increasing by a factor of $\approx 2 - 3$ at $z = 8 - 10$ relative to the $z = 6$ values discussed in Sect. \ref{sec:ratio}.

\section{[CII] luminosity}
\label{app:CII}

\begin{figure}
    \centering
    \includegraphics[width=0.9\columnwidth]{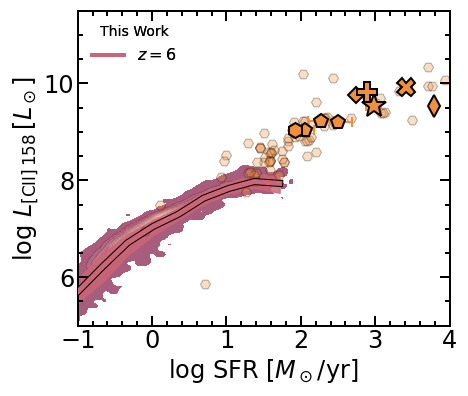}
    \caption{\small Relation between $\CII$ luminosity and SFR. Our simuation results at $z=6$ are shown as red contours (solid line indicating the median trend). Observational data at $z\simeq 6$ compiled by \cite{RP23} are shown as hexagons. We also include the high$-z$ sources with $\OI$ and/or $\OIdue$ detections shown in the main text: REBELS galaxies (pentagons), A1689-zD1 (hexagon), J2054-0005 (star), J2310+1855 (cross), W2246-0526 (plus), J204314$-$214439 (diamond) and SPT0346$-$52 (small diamond).}
    \label{fig:CIIsfr}
\end{figure}

In this Appendix we show our model's prediction for the [CII]-SFR relation in $z\sim 6$ galaxies. Given the larger abundance of [CII] data at this redshift with respect to [OI], we use this line as a sort of check of our line emission modeling. Also, we need to check the reliability of the [CII] predictions since we discuss some line ratios with this line. The results are reported in Fig. \ref{fig:CIIsfr}, where our theoretical predictions are compared with $z\sim 6$ observations from the compilation of \cite{RP23} and the objects with $\OI$ and/or $\OIdue$ detections adoped for comparison in this work (Sect. \ref{sec:obs}). The agreement up to the SFRs available in our simulation is good. In the high SFR end of our predictions there is a flattening of [CII] luminosity, due to the increasing abundance of CO and colder particles, which is at the origin of the well known [CII]/FIR deficit \citep[e.g.][]{Narayanan2017}. Such a prediction slightly deviates our model from the observations. However, we remark that a larger volume is needed to prove this high SFR regime.


\end{document}